\documentclass[prd,aps,preprint,showpacs,nofootinbib,superscriptaddress]{revtex4-1}
\usepackage{amsmath}
\usepackage{caption}
\usepackage{bm} 
  \usepackage{rotating}
  \usepackage{floatpag}
  \usepackage{subfig}
  \usepackage{mathrsfs}
  \rotfloatpagestyle{empty}
  \usepackage{bbold}

  \usepackage{amssymb}

  \usepackage{graphicx}

\usepackage{color}
\usepackage[normalem]{ulem}

\usepackage{lipsum}

\definecolor{myGreen}{rgb}{0.2,0.72,0.2}
\definecolor{myWhite}{rgb}{0.98,0.98,0.98}
\renewcommand\sout{\bgroup \color[rgb]{0.55,0.00,0.99} \ULdepth=-.5ex \ULset}

\begin{document}
\newcommand{\myfrac}[2]{\frac{#1}{#2}}
\newcommand{\lambdamb}{- {\textstyle \myfrac{1}{2}} \alpha}
\newcommand{\ket}[1]{| {#1} \rangle}
\newcommand{\bra}[1]{\langle {#1} |}
\newcommand{\braket}[2]{\langle {#1} | {#2} \rangle}
\newcommand{\norm}[1]{\| {#1} \|}
\newcommand{\diff}{{\rm d}}

\title{Collapse dynamics and Hilbert-space stochastic processes\footnote[0]{Corresponding author email: simone.rodini@unipv.it \\ Our colleague and friend Alberto sadly passed away before the completion of the present work; he lives  in our hearts.}
}


\author{Daniele Bajoni}
\affiliation{Dipartimento di Ingegneria Industriale e dell'Informazione - Universit\`a di Pavia - via A. Ferrata, 27100, Pavia,  Italy}
\author{Oreste Nicrosini}
\affiliation{Istituto Nazionale di Fisica Nucleare - Sezione di Pavia - via A. Bassi, 27100, Pavia, Italy} 
\affiliation{Dipartimento di Fisica - Universit\`a di Pavia - via A. Bassi, 27100, Pavia, Italy} 
\author{Alberto Rimini }
\affiliation{Istituto Nazionale di Fisica Nucleare - Sezione di Pavia - via A. Bassi, 27100, Pavia, Italy} 
\affiliation{Dipartimento di Fisica - Universit\`a di Pavia - via A. Bassi, 27100, Pavia, Italy}
\author{Simone Rodini}
\affiliation{Istituto Nazionale di Fisica Nucleare - Sezione di Pavia - via A. Bassi, 27100, Pavia, Italy} 
\affiliation{Dipartimento di Fisica - Universit\`a di Pavia - via A. Bassi, 27100, Pavia, Italy}

\begin{abstract}
Spontaneous collapse models of state vector reduction represent a possible solution to the 
quantum measurement problem.  In the present paper we focus our attention on the Ghirardi-Rimini-Weber 
(GRW) theory and the corresponding continuous localisation models in the form of a Brownian-driven motion in Hilbert space. 
We consider experimental setups in which a single photon hits a beam splitter and is subsequently 
detected by photon detector(s), generating a superposition of photon-detector quantum states. 
Through a numerical approach we study the dependence of collapse times on the physical features 
of the superposition generated, including also the effect of a finite reaction time of the measuring apparatus. 
We find that collapse dynamics is sensitive to the number of detectors and the physical properties of the 
photon-detector quantum states superposition.
\end{abstract}
\maketitle
\section{Introduction}
\label{intro}

A number of interpretations/extensions of quantum mechanics have been proposed to attempt to solve the quantum measurement problem.
On the one hand, there, for instance, are decoherence theories: see ref.~\cite{Schlosshauer:2005} for a review. On the other hand, various extensions of quantum mechanics have been developed that lead to dynamical models for the collapse of the wave function. Stochastic processes modelling the spontaneous collapse of the state of the quantum system were first proposed in ref.~\cite{Ghirardi:1985mt}. This is the so called GRW theory, which involves discontinuous stochastic processes \cite{ShanGao2018,Wechsler:2020}.  Continuous models have been also been devised, in which the spontaneous collapse of the quantum state is realized in the form of a continuous stochastic process in Hilbert space. A number of different versions of such models have been put forward. See, for example, refs.~\cite{Ghirardi:1989cn,Bassi:2012bg,Diosi:1988,Ghirardi:1989cn,Ghirardi:1985mt,Gisin:1984,Adler_2001,Adler_2002,Adler_2003,Brody_2002,Di_si_1988,Gisin_e1,Hughston_e1,Penrose_e1,Percival_e1}. 

 The two families of processes are strictly connected. Actually, in ref.~\cite{NicrosiniRimini:1990} it has been shown that discontinuous processes, in a proper infinite frequency limit, are equivalent to appropriate continuous ones. 
 In all these process in  Hilbert space a set of {\it physical quantities} (observables) appears, 
represented by  the corresponding set of selfadjoint operators. 
The  processes act inducing the {\it sharpening } of the  distribution of values of those quantities around a 
stochastically chosen centre. In ref.~\cite{ShanGao2018} it has been shown that the physical effect of the stochastic processes depends on the choice of observables that are being {\it sharpened} and not on the details of the sharpening procedure. 

In this work we focus the attention on the position of a macroscopic/mesoscopic pointer as the observable that is going to undergo the sharpening processes. In particular, we examine two types of experimental setups, in which a superposition of macro(meso)scopic states is generated. By using a continuous model for the collapse, we show how the collapse times depends both on the number of detectors and, crucially, on the physical features of the superposition in the measured state.  By using the connection between the continuous process and the GRW model, we provide also a clear physical interpretation of the parameters of the model. This type of studies represent an essential step to guide actual experimental effort, in order to be able to impose limits on the values of the parameters and either confirm or disprove the spontaneous collapse models.

The paper is organised as follows. In Sect.~\ref{sect:themodels} the main features of the two approaches are 
recalled, together with their relationship. In Sect.~\ref{sect:onemeasapp} the continuous process is specialised 
to an experimental setup in which a single photon is sent to a beam splitter creating a superposition of 
transmitted/reflected photon states, and the photon is either detected  or not detected by a single-photon detector placed 
in the transmission region. Sect.~\ref{sect:twomeasapp}  is devoted to the analysis of the setup in 
which a second single-photon detector is added in the reflection region. In Sect.~\ref{sect:detmod} the photon/detector interaction 
is modelled and taken into account in the measurement dynamics, together with its interplay with the stochastic reduction process. 
In Sect.~\ref{sec:numres} a number numerical simulations are shown and commented. In Sect.~\ref{sect:concl} some conclusions are drawn. 

\section{Hitting  and continuous processes}
\label{sect:themodels}

Let the set of compatible quantities characterizing the discontinuous stochastic process be 
\begin{equation}
{\hat{\bm A}} \equiv \{ \hat A_m; \ m=1,2,\dots,K\}, \qquad  \big[\hat A_m\,,\;\hat A_n\big]=0, \qquad \hat{A}^\dagger_m = \hat{A}_m , 
\end{equation}
and the sharpening action be given by the operator
\begin{equation}
S_i = \left(\myfrac{\alpha}{\pi}\right)^{\;K\;/ \;4} \; \exp(R_i) , \qquad R_i= \lambdamb \,(\hat{\bm A} -\bm a_i)^2 . 
\end{equation}
The parameter $ \alpha $ rules the accuracy of the sharpening and $\bm a_i $ is the centre of the $i$-th hitting. 
It is assumed that the hittings occur randomly in time, distributed according to a Poisson law with frequency $\lambda$. 

The sharpening operator for the $i$-th hitting $S_i$ acts on the normalized state vector $\ket{\psi_t}$ giving the state vector $\ket{\varphi_{i,t}}$, which can be recast in a normalized vector $\ket{\psi'_{i,t}}$:\begin{equation}
\label{eq:singlehitting}
\ket{\psi'_{i,t}} = \myfrac{\ket{\varphi_{i,t}}}{\norm{\varphi_{i,t}}} , 
\hskip50pt
\ket{\varphi_{i,t}} = S_i \ket{\psi_{t}} .  
\end{equation}
The probability that the hitting takes place around $\bm a_i $ is 
 \begin{equation}
 \label{eq:singleprob}
{\mathscr P} \left( \psi_t \vert \bm a_i  \right) = {\norm{\varphi_{i,t}}}^2 .
\end{equation}

Actually, it turns out that the effectiveness of the discontinuous process 
depends on $\alpha$ and $\lambda$ only through their product. 

The continuous process based on the same quantities ${\hat{\bm A}} \equiv \{ \hat A_m\}_{m=1}^K$ is ruled by 
the It{\^ o} stochastic differential equation 
\begin{equation}
\diff\ket{\psi} 
= \Big[\sqrt{\gamma}\big(\hat{\bm A} \;-\;\langle\hat{\bm A}\rangle_{\;\psi_t}\big) \cdot \diff\bm B
-{\textstyle\myfrac{1}{2}} \gamma \big(\hat{\bm A} \;-\;\langle\hat{\bm A}\rangle_{\;\psi_t}\big)^2 \diff t \Big]
\,\ket\psi , 
\label{eq:genstochproc}
\end{equation}
where
\begin{equation}
\diff{\bm B}\equiv\{\diff B_m;\ m=1,2,\dots,K\}, \qquad \overline{\diff\bm B} = 0 , \qquad \overline{\diff B_m\,\diff B_n} = 
\delta_{mn} \,\diff t, 
\label{eq:genBstat}
\end{equation}
and 
\begin{equation}
\langle\hat{\bm A}\rangle_{\;\psi_t} = \bra{\psi_t } \hat{\bm A} \ket{\psi_t } . 
\label{eq:genAav}
\end{equation}
The parameter $\gamma$ sets the effectiveness of the process. 
 
In refs.~\cite{NicrosiniRimini:1990} and \cite{ShanGao2018} it has been shown 
 that, by taking the infinite frequency limit of the discontinuous process (\ref{eq:singlehitting}) and (\ref{eq:singleprob}) with the prescription 
\begin{equation}
\label{eq:betamu}
\alpha \lambda = {\rm constant} = 2 \gamma, 
\end{equation}
one gets the continuous process of eq.~(\ref{eq:genstochproc}). As a consequence it becomes apparent that, for $t \to \infty$, the continuous process drives the state vector to a common eigenvector of the operators $\hat {\bm A}$.
The probability of a particular eigenvector $\ket{\bm a_r}$ is given by 
$\vert \braket{\bm a_r}{\psi_0} \vert^2$, for the generic state vector $\ket{\psi_0}$  at a given arbitrary initial time.

\section{Single detector}
\label{sect:onemeasapp}

\begin{figure}[h]
\begin{center}
\includegraphics[width=0.7\textwidth]{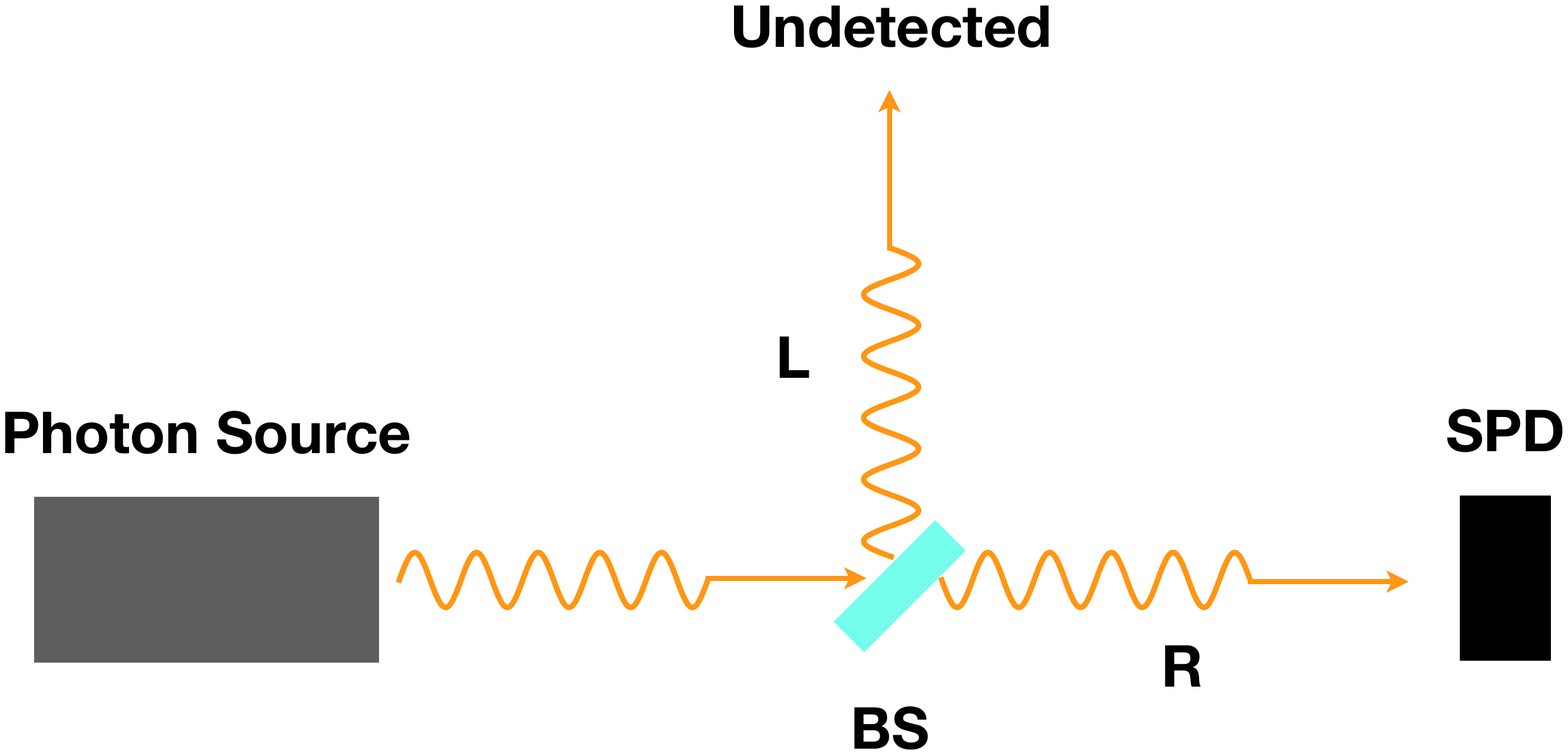}
\end{center}
\caption{Schematic representation of a single detector setup for generating and measuring superposition state. A single photon is emitted from the source and hits a beam splitter (BS), generating a superposition of $\ket{\gamma}_R$ and $\ket{\gamma}_L$, namely $\ket{\psi} = c_L\ket{\gamma}_L+c_R\ket{\gamma}_R$. A single photon detector (SPD) is placed in $R$ position.}
\label{fig:exp1}
\end{figure}

Consider the experimental setup depicted in Figure~\ref{fig:exp1}, in which a single photon state hits a beam splitter (BS), a superposition of $\ket{\gamma}_R$ and $\ket{\gamma}_L$ states is formed and a single photon detector (SPD) is placed in $R$ position. 
In this case eqs.~(\ref{eq:genstochproc})--(\ref{eq:genAav}) are specialized to  a single operator $\hat A$, namely the operator associated to the ``pointer position" (more on this later). 

We indicate with $\vert c_{R,L}\vert^2$ the transmission and reflection coefficients of the BS, respectively. Therefore, in the {\it photon + detector} Hilbert space one has that the Schr\"odinger evolution generates the superposition

\begin{eqnarray}
\ket{\psi} = \Big( c_L \ket{ \gamma}_L +  c_R \ket{ \gamma}_R \Big) \ket{D^{(0)}} &\rightarrow&  
c_L \ket{ \gamma}_L  \ket{D^{(0)}}  
+ c_R \ket{ \gamma}_R \ket{D^{(+)}} \nonumber \\
&=& c_L \ket{\psi_L} + c_R \ket{\psi_R} , 
\label{eq:super1}
\end{eqnarray}
where $\ket{D^{(0,+)}} $ are the SPD states that correspond to SPD ``ready" or ``clicked", respectively, 
and  $\vert c_L \vert^2 + \vert c_R \vert^2 = 1$. 
The operator associated to the pointer position can be represented as 
\begin{equation}
\hat A =  \mathbb{1}_\gamma \otimes \left[ a_0 \ket{D^{(0)}} \bra{D^{(0)}} + a_+ \ket{D^{(+)}} \bra{D^{(+)}}  \right] , 
\end{equation}
where $a_{0,+}$ are the pointer position eigenvalues. Eqs.~(\ref{eq:genstochproc})--(\ref{eq:genAav}) become 
\begin{align}
\diff\ket{\psi} 
&= \Big[\sqrt{\gamma}\big(\hat{ A} \; - \; \langle\hat{ A}\rangle_{\;\psi_t}\big)  \diff B
-{\textstyle\myfrac{1}{2}} \gamma \big(\hat{ A} \; - \; \langle\hat{A}\rangle_{\;\psi_t}\big)^2 \diff t \Big]
\,\ket\psi , \label{eq:contproc1}\\
\overline{\diff B} &= 0 , \qquad \overline{(\diff B)^2} = \diff t , 
 \label{eq:1dimdbstat}\\
  \langle\hat{ A}\rangle_{\;\psi_t} &= a_0 \vert c_L  \vert^2 + a_+ \vert c_R  \vert^2 = J(t) . 
\label{eq:1dimAav}
\end{align}
The time evolution of the coefficients of the superposition, $c_{R,L}$, as due to the stochastic process, can be obtained by projecting 
$ \ket{\psi(t +\diff t)}$ onto  $\ket{\psi_{R,L}}$, namely
\begin{equation}
\label{eq:ctpdtod}
c_{R,L} (t + \diff t) = \braket{\psi_{R,L}} {\psi(t +\diff t) } = \bra{\psi_{R,L}} \Big( \ket{\psi(t)} + \diff \ket{\psi} \Big) = L_{R,L} (t) c_{R,L} (t) , 
\end{equation}
where 
\begin{equation}
L_{R,L} (t) = 1 + \Big[  \sqrt{\gamma} K_{+,0} (t) \diff B - {\textstyle\myfrac{1}{2}} \gamma K_{+,0}^2(t) \diff t\Big] ,
\end{equation}
and $K_{0,+} (t) = a_{0,+} - J(t)$. 
Since $K_i(t)$ (and hence $L_i(t)$) is invariant under translation of the system of eigenvalues $a_i$, without loss of generality one can take 
$a_0 = 0 $ and $a_+ = a$, obtaining
\begin{align}
K_0(t) &=  -a \vert c_R \vert^2 , \\ 
K_+(t) &= a \Big(  1 - \vert c_R \vert^2 \Big) = a \vert c_L \vert^2 , 
\end{align}
from which 
\begin{equation}
\label{eq:onedetcoeef}
L_{R,L} (t) =  1 + \Big[  \pm a \sqrt{\gamma} \vert c_{L,R}(t) \vert^2  \diff B - {\textstyle\myfrac{1}{2}} \gamma a^2 \vert c_{L,R} (t) \vert^4  \diff t \Big]  . 
\end{equation}
Since, according to eq.~(\ref{eq:ctpdtod}), 
\begin{equation}
\vert c_{R,L} (t + \diff t) \vert^2 =  L^2_{R,L} (t) \vert c_{R,L} (t) \vert^2 , 
\label{eq:tevolc}
\end{equation}
one has, by squaring  eq.~(\ref{eq:onedetcoeef}) and taking into account the It{\^o} lemma, that:
\begin{equation}
L^2_{R,L} (t) = 1 \pm 2 a \sqrt{\gamma} \vert c_{L,R}(t) \vert^2  \diff B . 
\label{eq:evolfact}
\end{equation}
The stochastic factor $L^2_{R,L} (t)$ of eq.~(\ref{eq:evolfact}) can be generated numerically in each step in $dt$ by extracting  
a gaussian random number $\diff B$, according to the statistics of eq.~(\ref{eq:1dimdbstat}), 
and iteratively inserting it into eq.~(\ref{eq:tevolc})
 to produce the ``path" followed by $\vert c_{R,L} \vert^2$ during the reduction process. 
\section{Two detectors}
\label{sect:twomeasapp}
\begin{figure}[h]
\begin{center}
\includegraphics[width=0.7\textwidth]{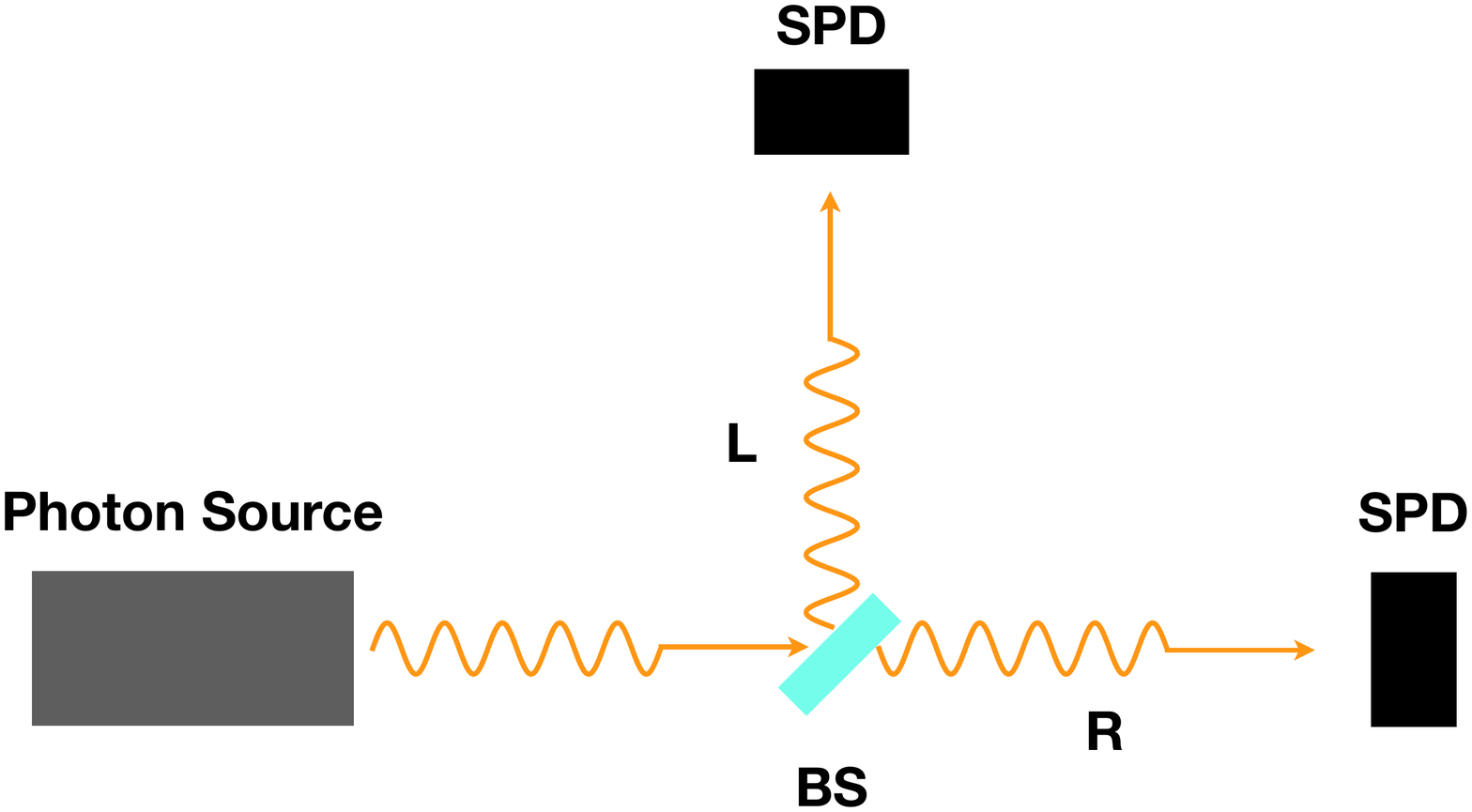}
\end{center}
\caption{Schematic representation of a double detector setup for generating and measuring superposition state. A single photon is emitted from the source and hits a beam splitter (BS), generating a superposition of $\ket{\gamma}_R$ and $\ket{\gamma}_L$, namely $\ket{\psi} = c_L\ket{\gamma}_L+c_R\ket{\gamma}_R$. A pair of single photon detectors (SPD) are placed in $R$ and $L$ positions.}
\label{fig:exp2}
\end{figure}
\noindent
When considering the experimental setup described in Figure~\ref{fig:exp2}, where a second (and, for the sake of simplicity, identical) SPD has been added in $L$ 
position, the Schr\"odinger evolution generates, in the {\it photon + detectors} Hilbert space, the superposition 
\begin{align}
\ket{\psi} &= \Big( c_L \ket{ \gamma}_L +  c_R \ket{ \gamma}_R \Big) \ket{D_L^{(0)}} \ket{D_R^{(0)}} \nonumber \\
& \rightarrow
c_L \ket{ \gamma}_L  \ket{D_L^{(+)}}   \ket{D_R^{(0)}}  
+ c_R \ket{ \gamma}_R \ket{D_L^{(0)}}  \ket{D_R^{(+)}}
= c_L \ket{\psi_L} + c_R \ket{\psi_R} , 
\label{eq:super2}
\end{align}
where now $D_{R,L}$ are the states of the $R$ and $L$ SPD, respectively, and, again, $\vert c_L \vert^2 + \vert c_R \vert^2 = 1$. 
The operators associated to the $R$ and $L$ pointer position are 
\begin{equation}
\hat A_{R,L}=  \mathbb{1}_\gamma \otimes \mathbb{1}_{L,R} \otimes \left[ a_0 \ket{D^{(0)}_{R,L}} \bra{D^{(0)}_{R,L}} 
+ a_+ \ket{D^{(+)}_{R,L}}  \bra{D^{(+)}_{R,L}}  \right]
\end{equation}
and  eqs.~(\ref{eq:contproc1})--(\ref{eq:1dimAav}) become 
\begin{align}
\diff\ket{\psi} 
\label{eq:contproc2}
&=\Big[ \sqrt{\gamma} \big(\hat{ A_L} \; - \; \langle\hat{ A_L}\rangle_{\;\psi_t}\big)  \diff B_L
-{\textstyle\myfrac{1}{2}} \gamma \big(\hat{ A_L} \; - \; \langle\hat{A_L}\rangle_{\;\psi_t}\big)^2 \diff t \nonumber \\
 & + \sqrt{\gamma} \big(\hat{ A_R} \; - \; \langle\hat{ A_R}\rangle_{\;\psi_t}\big)  \diff B_R
-{\textstyle\myfrac{1}{2}} \gamma \big(\hat{ A_R} \; - \; \langle\hat{A_R}\rangle_{\;\psi_t}\big)^2 \diff t \Big] \,\ket\psi , \\
 \overline{\diff B_{R,L}} &= 0 , \qquad \overline{(\diff B_{R,L})^2} = \diff t , \qquad \overline{\diff B_{R} \diff B_{L}} = 0 , \\
\langle\hat{ A_R}\rangle_{\;\psi_t} &= \langle\hat{ A_L}\rangle_{\;\psi_t} = \langle\hat{ A}\rangle_{\;\psi_t} = a_0 \vert c_L  \vert^2 + a_+ \vert c_R  \vert^2 . 
\end{align}
By following the same procedure sketched in the previous section, one obtains
\begin{align}
\label{eq:twodetcoeff}
L_{R,L} (t) =  1 + \Big[  &\pm a \sqrt{\gamma} \vert c_{L,R} \vert^2  \diff B_{R,L} - {\textstyle\myfrac{1}{2}} \gamma a^2 \vert c_{L,R} \vert^4  \diff t  \nonumber \\
&\mp a \sqrt{\gamma} \vert c_{L,R} \vert^2  \diff B_{L,R} - {\textstyle\myfrac{1}{2}} \gamma a^2 \vert c_{L,R} \vert^4  \diff t  \Big] , 
\end{align}
from which
\begin{equation}
L^2_{R,L} (t) =  1   \pm 2 a \sqrt{\gamma} \vert c_{L,R} \vert^2  \diff B_{R,L} \mp 2 a \sqrt{\gamma} \vert c_{L,R} \vert^2  \diff B_{L,R}  . 
\label{eq:stochfact2}
\end{equation}
It is worth noticing that eq.~(\ref{eq:stochfact2}) can be rewritten as 
\begin{equation}
L^2_{R,L} (t) =  1   \pm 2 a \sqrt{\gamma} \vert c_{L,R} \vert^2  \left(  \diff B_{R} -  \diff B_{L} \right) = 
1   \pm 2 a \sqrt{\gamma} \vert c_{L,R} \vert^2 {\diff C} , 
\label{eq:tdevolfact}
\end{equation}
where 
\begin{equation}
{\diff C} =  \diff B_{R} -  \diff B_{L} 
\end{equation}
is a gaussian stochastic variable obeying the statistics
\begin{equation}
\overline{\diff C} = 0 , \quad \overline{(\diff C)^2} =  \overline{(\diff B_R)^2} + \overline{(\diff B_L)^2} = 2 \diff t . 
\end{equation}

\section{The system/detector interaction and its interplay with the stochastic process}
\label{sect:detmod}
In the previous sections the system/apparatus interactions generating the superpositions (\ref{eq:super1}) and (\ref{eq:super2}) have been considered ``instantaneous". Actually, any real device has a finite reaction time $T$ and the dynamical development of the superpositions can be modelled by proper interaction hamiltonians (see for instance ref.~\cite{Mello2010}, where the 
von Neumann model of measurement in quantum mechanics is described in detail, and ref.~\cite{Adler:2020} for a recent discussion about the interplay between the measurement time and the localisation process). 

Considering the single detector case, the system/apparatus interactions hamiltonian can be written as 
\begin{equation}
\label{eq:SAinteractionH}
 \hat H_I = \myfrac{\diff \beta}{\diff t} f(\hat R) \hat P , 
\end{equation}
where $\hat P$ is the momentum operator of the pointer, $R$ is defined as 
\begin{equation}
\hat{R} |\gamma\rangle_R = |\gamma\rangle_R \qquad  \hat{R} |\gamma\rangle_L = 0 , 
\end{equation}
the function $f$ satisfies 
\begin{equation}
f(0)  = 0,  \qquad  f(1) = a , 
\end{equation}
 $\beta(t)$ is the activation function that, for the sake of simplicity, is assumed piecewise linear starting from 
$t=0$ taken as the interaction initial time 
\begin{equation}
\beta(t) = {\rm min} (t/T, 1) \qquad t\geq 0 , 
\end{equation}
and $T$ is the time at which the interaction has fully developed.  

The effect of the interaction hamiltonian (\ref{eq:SAinteractionH}) on the initial superposition 
\begin{equation}
\ket{\psi} = \Big( c_L \ket{ \gamma}_L +  c_R \ket{ \gamma}_R \Big) \ket{D^{(0)}} 
\end{equation}
is the following: since $\hat R \ket{\gamma}_L = 0 $, the term $c_L \ket{ \gamma}_L  \ket{D^{(0)}} $ is left unchanged; on the contrary, 
since $\hat R \ket{\gamma}_R = \ket{\gamma}_R$, the pointer position in the term $c_R \ket{ \gamma}_R  \ket{D^{(0)}} $ is shifted 
from position ``0" to position ``$a$", corresponding to $c_R \ket{ \gamma}_R  \ket{D^{(+)}} $, linearly in the time interval $(t=0, t=T)$. 

When taking into account the interaction (\ref{eq:SAinteractionH}),  eq.~(\ref{eq:contproc1}) becomes 
\begin{equation}
\label{eq:contproc1+SAint}
\diff\ket{\psi} 
= \Big[ - \myfrac{i}{\hbar} \hat H_I \diff t + \sqrt{\gamma}\big(\hat{ A} \; - \; \langle\hat{ A}\rangle_{\;\psi_t}\big)  \diff B
-{\textstyle\myfrac{1}{2}} \gamma \big(\hat{ A} \; - \; \langle\hat{A}\rangle_{\;\psi_t}\big)^2 \diff t \Big]
\,\ket\psi . 
\end{equation}

\begin{figure}[htb]
\centering
\subfloat[][Starting condition: $|c_R|^2 = \frac{1}{2}$.]{\includegraphics[width=0.45\textwidth]{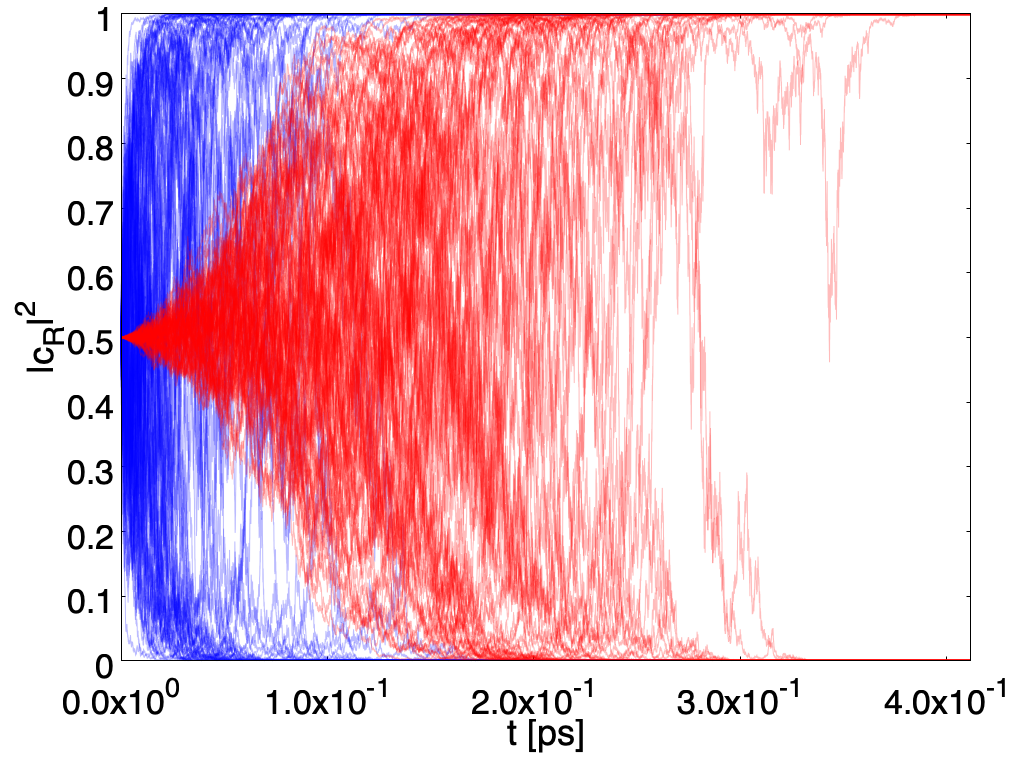}\label{oneDetector_m1_a}}
\subfloat[][Starting condition: $|c_R|^2 = \frac{2}{3}$.]{\includegraphics[width=0.45\textwidth]{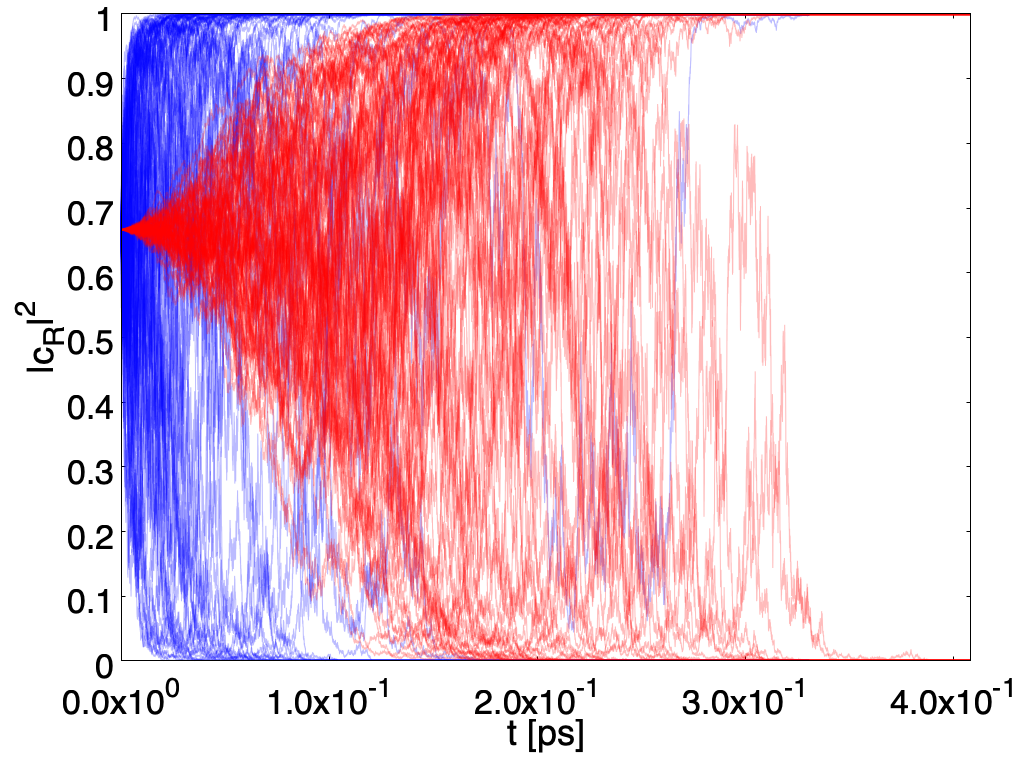}\label{oneDetector_m1_b}} \\
\subfloat[][Starting condition: $|c_R|^2 = \frac{4}{5}$.]{\includegraphics[width=0.45\textwidth]{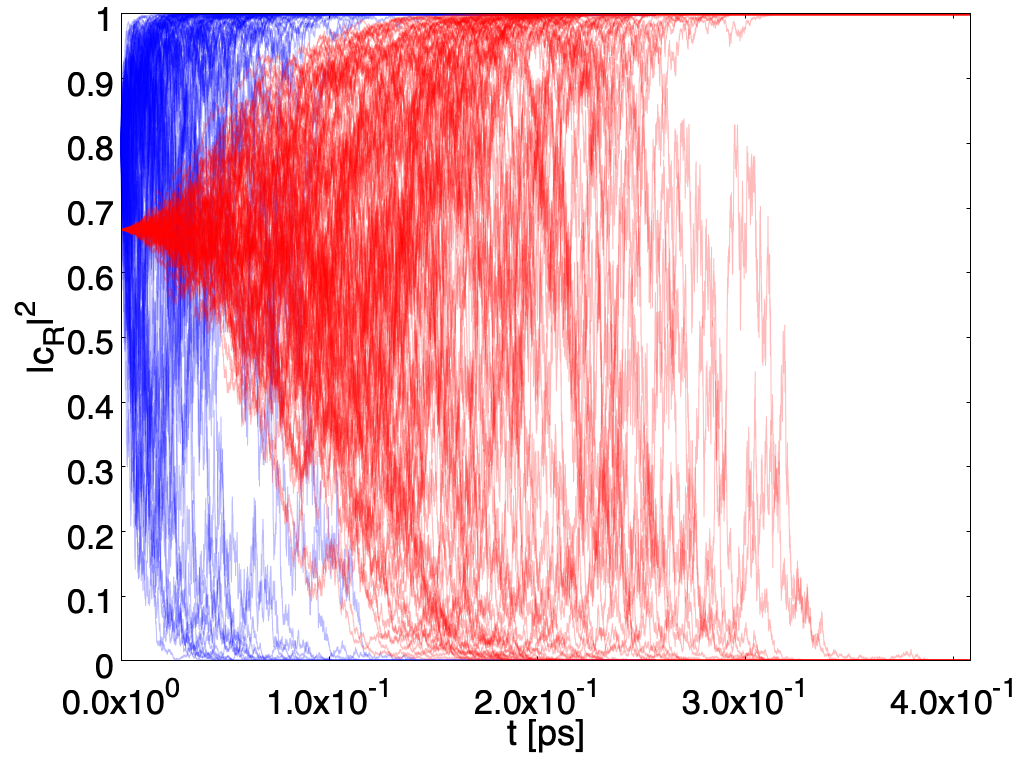}\label{oneDetector_m1_c}}
\subfloat[][Starting condition: $|c_R|^2 = \frac{1}{20}$.]{\includegraphics[width=0.45\textwidth]{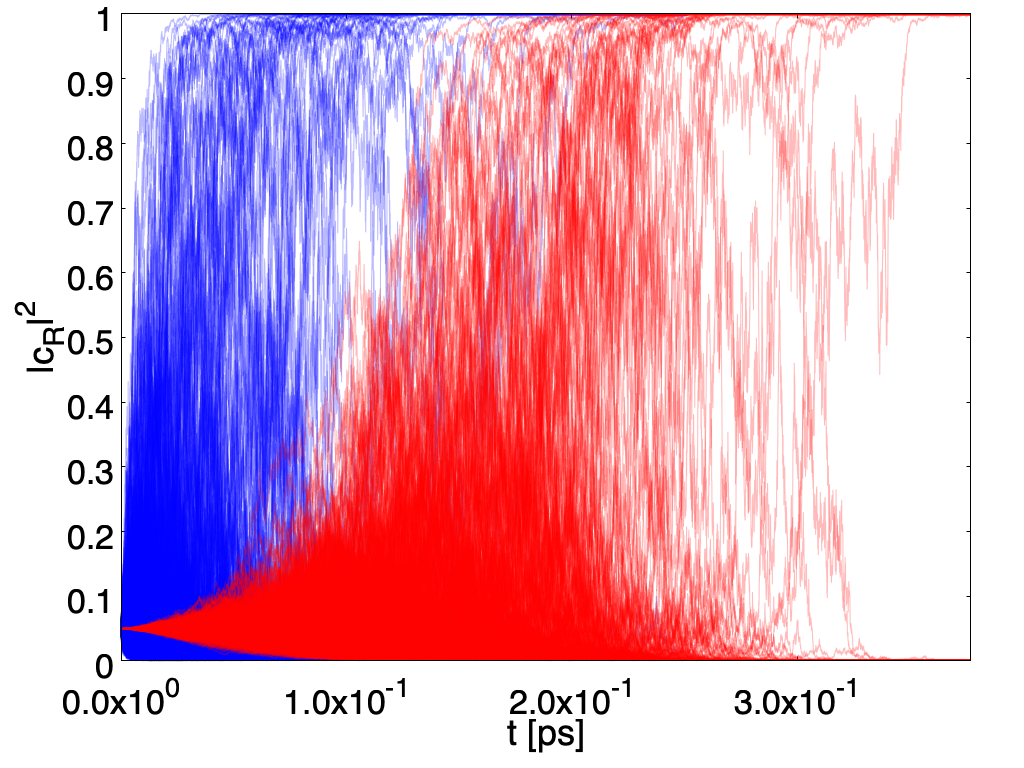}\label{oneDetector_m1_d}}
\caption{$N=1000$ paths of the probability of having the single photon state in the $R$ sector ($|c_R|^2$) as function of $t$ in case of a single photon detector with $\gamma = \gamma_1$ (see also Figure \ref{fig:exp1}). The time scale in ns$=10^{-9}$ s. For all the figures, in blue are the paths for the case without the von Neumann activation   delay for the detector, in red are the paths with  von Neumann activation for $T=10^{-4}$ ns.}
\label{oneDetector_m1}
\end{figure}

Under the (very good) approximation that the stochastic process does not affect the pointer wave functions corresponding to 
different pointer positions, but affects only the coefficients of the superposition ($c_{R,L}$), eq.~(\ref{eq:onedetcoeef}) becomes 
\begin{equation}
\label{eq:onedetcoeefw}
L_{R,L} (t) =  1 + \Big[  \pm a \sqrt{\gamma} \beta(t) \vert c_{L,R} \vert^2  \diff B - {\textstyle\myfrac{1}{2}} \gamma a^2  \beta^2 (t) \vert c_{L,R} \vert^4  \diff t \Big] 
\end{equation}
from which 
\begin{equation}
\label{eq:onedetcoeefws}
L^2_{R,L} (t) =  1  \pm 2 a \sqrt{\gamma} \beta(t) \vert c_{L,R} \vert^2  \diff B 
\end{equation}
and, analogously, eq.~(\ref{eq:twodetcoeff}) turns into
\begin{eqnarray}
\label{eq:twodetcoeffw}
L_{R,L} (t) =  1 + \Big[  &\pm& a \sqrt{\gamma} \beta(t) \vert c_{L,R} \vert^2  \diff B_{R,L} - {\textstyle\myfrac{1}{2}} \gamma a^2 
\beta^2(t) \vert c_{L,R} \vert^4  \diff t  \nonumber \\
&\mp& a \sqrt{\gamma} \beta(t) \vert c_{L,R} \vert^2  \diff B_{L,R} - {\textstyle\myfrac{1}{2}} \gamma a^2 \beta^2(t) \vert c_{L,R} \vert^4  \diff t  \Big] 
\end{eqnarray}
from which
\begin{eqnarray}
\label{eq:twodetcoeffws}
L^2_{R,L} (t) &=&  1 \pm 2 a \sqrt{\gamma} \beta(t) \vert c_{L,R} \vert^2  \diff B_{R,L}
\mp 2 a \sqrt{\gamma} \beta(t) \vert c_{L,R} \vert^2  \diff B_{L,R}  \nonumber \\
&=& 1   \pm 2 a \sqrt{\gamma} \beta(t) \vert c_{L,R} \vert^2 {\diff C} . 
\label{eq:twodetcoeefws}
\end{eqnarray}

\begin{figure}[htb]
\centering
\subfloat[][Starting condition: $|c_R|^2 = \frac{1}{2}$.]{\includegraphics[width=0.45\textwidth]{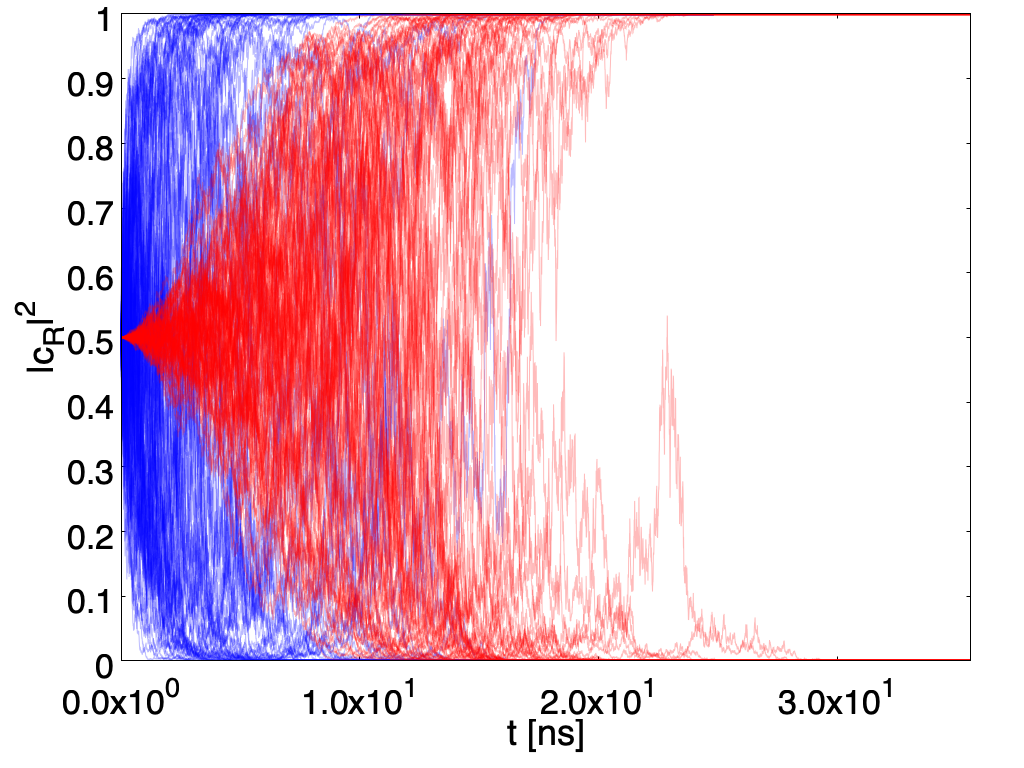}\label{oneDetector_m2_a}}
\subfloat[][Starting condition: $|c_R|^2 = \frac{2}{3}$.]{\includegraphics[width=0.45\textwidth]{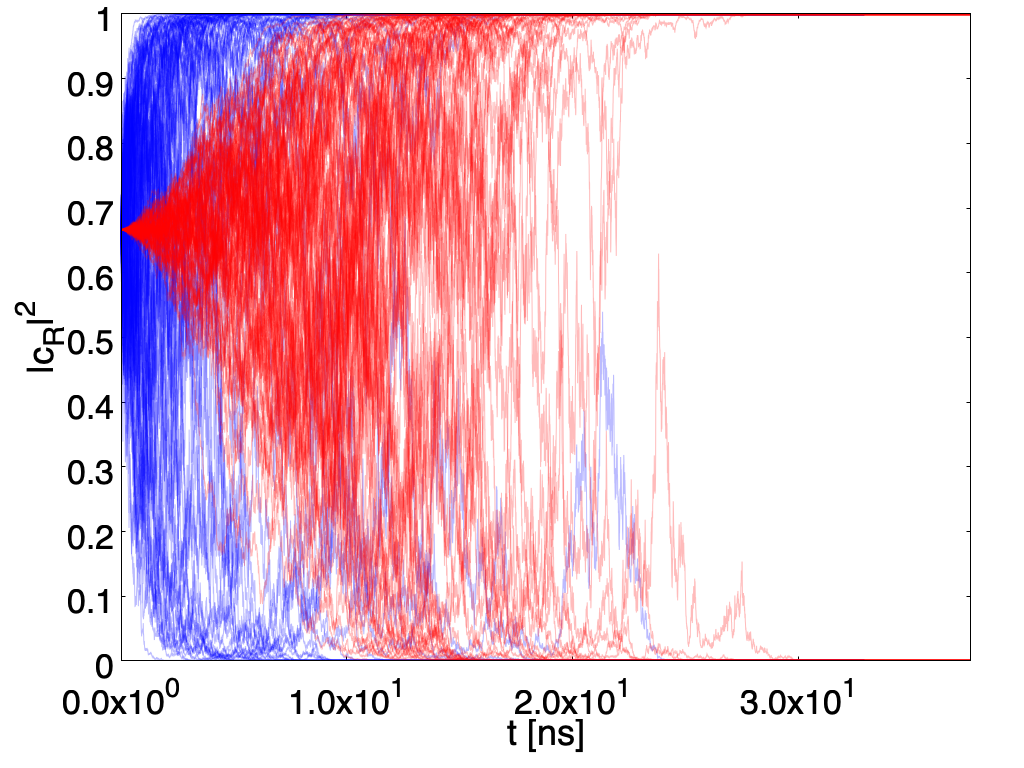}\label{oneDetector_m2_b}} \\
\subfloat[][Starting condition: $|c_R|^2 = \frac{4}{5}$.]{\includegraphics[width=0.45\textwidth]{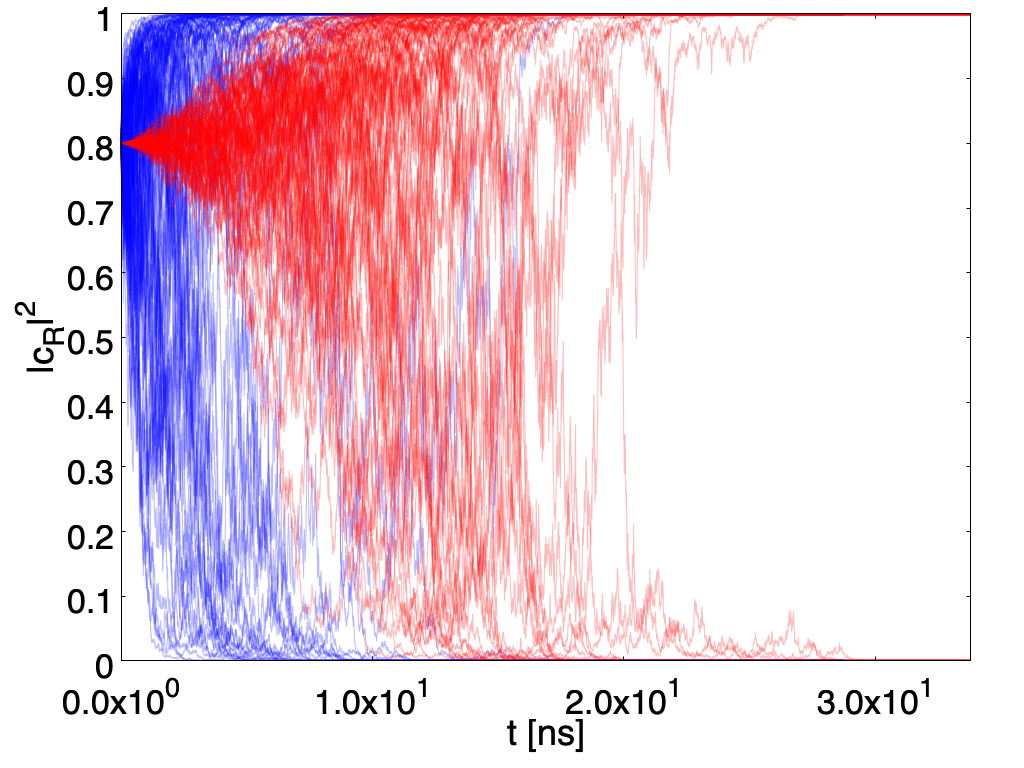}\label{oneDetector_m2_c}}
\subfloat[][Starting condition: $|c_R|^2 = \frac{1}{20}$.]{\includegraphics[width=0.45\textwidth]{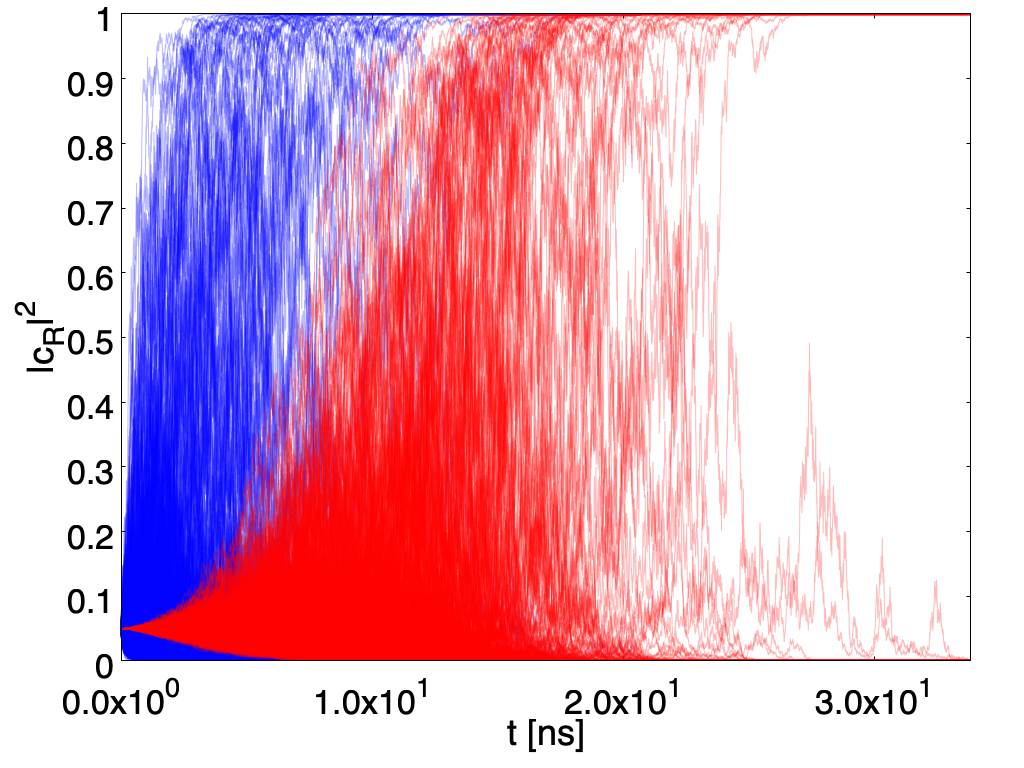}\label{oneDetector_m2_d}}
\caption{$N=1000$ paths of the probability of having the single photon state in the $R$ sector ($|c_R|^2$) as function of $t$ in case of a single detector with $\gamma = \gamma_2$. The time scale in ns$=10^{-9}$ s. For all the figures, in blue are the paths for the case without the von Neumann activation   delay for the detector, in red are the paths with  von Neumann activation for $ T =5$ ns. }
\label{oneDetector_m2}
\end{figure}

It is worth noticing that the difference between eqs.~(\ref{eq:evolfact}) and (\ref{eq:tdevolfact}) on one side, and 
eqs.~(\ref{eq:onedetcoeefws}) and (\ref{eq:twodetcoeefws}) on the other, consists in the presence of the activation function 
$\beta(t)$ in the stochastic term, and can be qualitatively understood  as follows: the stochastic localization process is the more 
effective the more separate are the positions of the terms in the superposition. Since the interaction hamiltonian (\ref{eq:SAinteractionH}) generates the superposition gradually, the stochastic process starts at time $t=0$ and becomes fully effective at time $t=T$, when $\beta(T) = 1$. 

\begin{figure}[htb]
\centering
\subfloat[][Starting condition: $|c_R|^2 = \frac{1}{2}$.]{\includegraphics[width=0.45\textwidth]{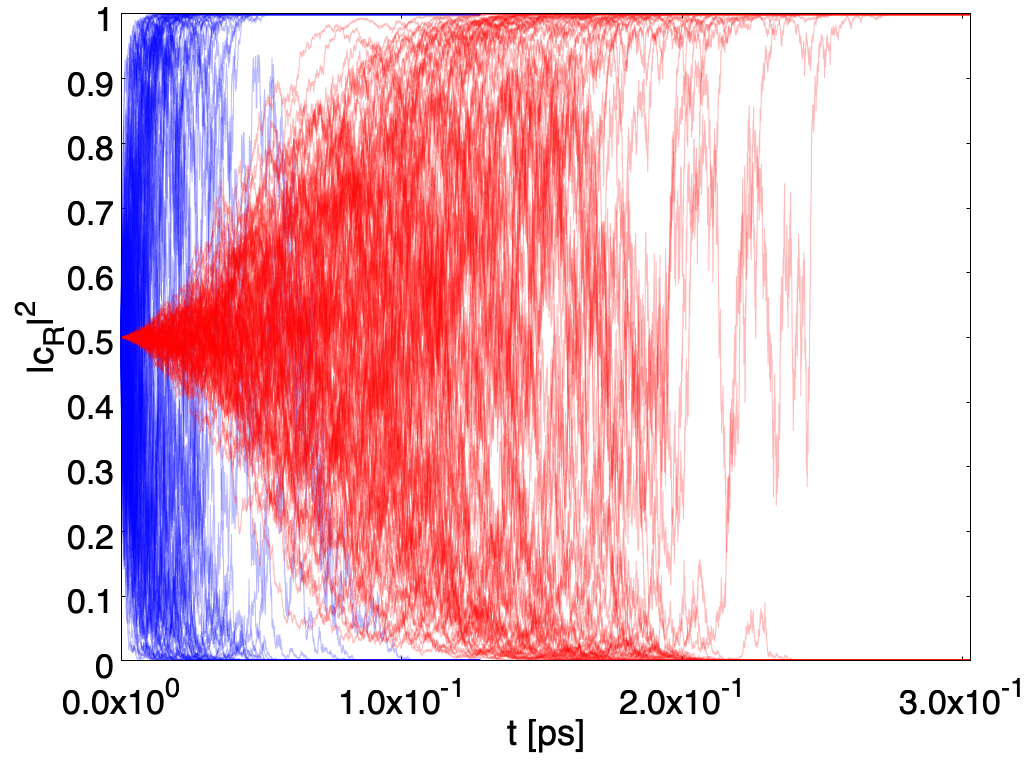}\label{twoDetector_m1_a}}
\subfloat[][Starting condition: $|c_R|^2 = \frac{2}{3}$.]{\includegraphics[width=0.45\textwidth]{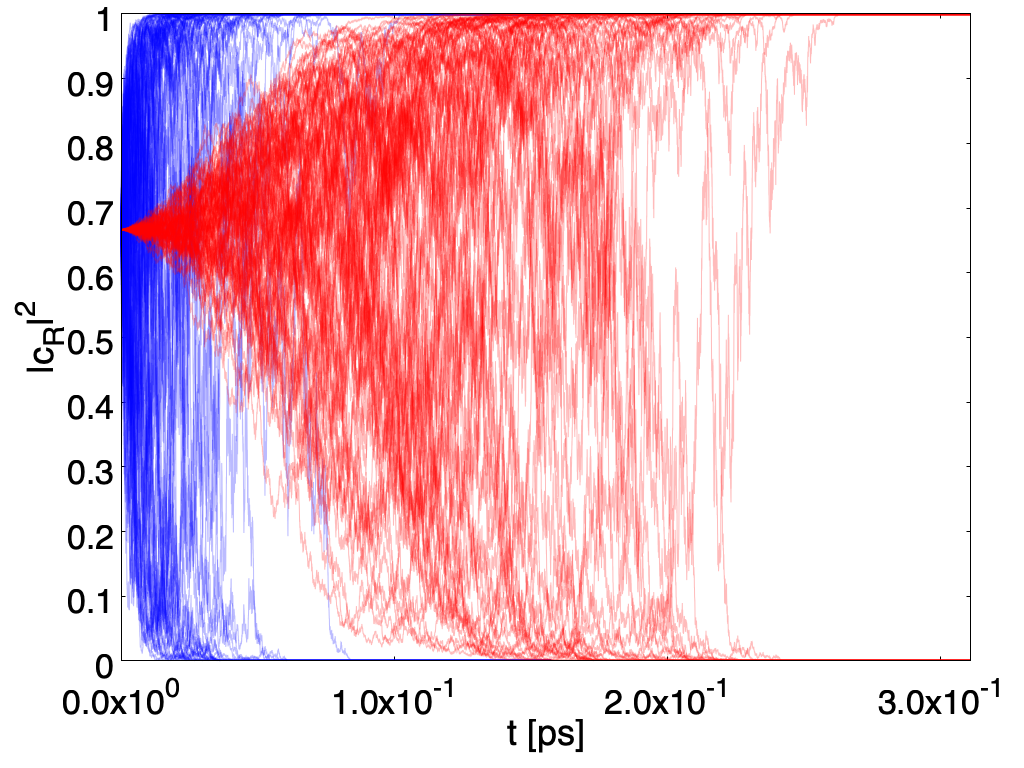}\label{twoDetector_m1_b}} \\
\subfloat[][Starting condition: $|c_R|^2 = \frac{4}{5}$.]{\includegraphics[width=0.45\textwidth]{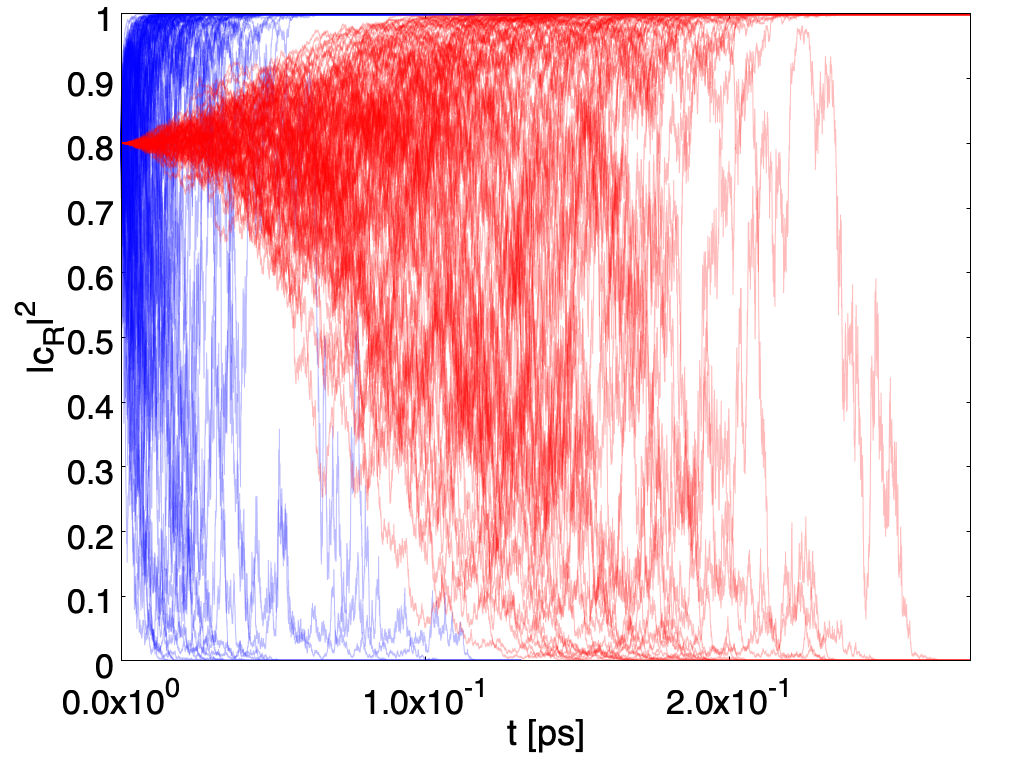}\label{twoDetector_m1_c}}
\subfloat[][Starting condition: $|c_R|^2 = \frac{1}{20}$.]{\includegraphics[width=0.45\textwidth]{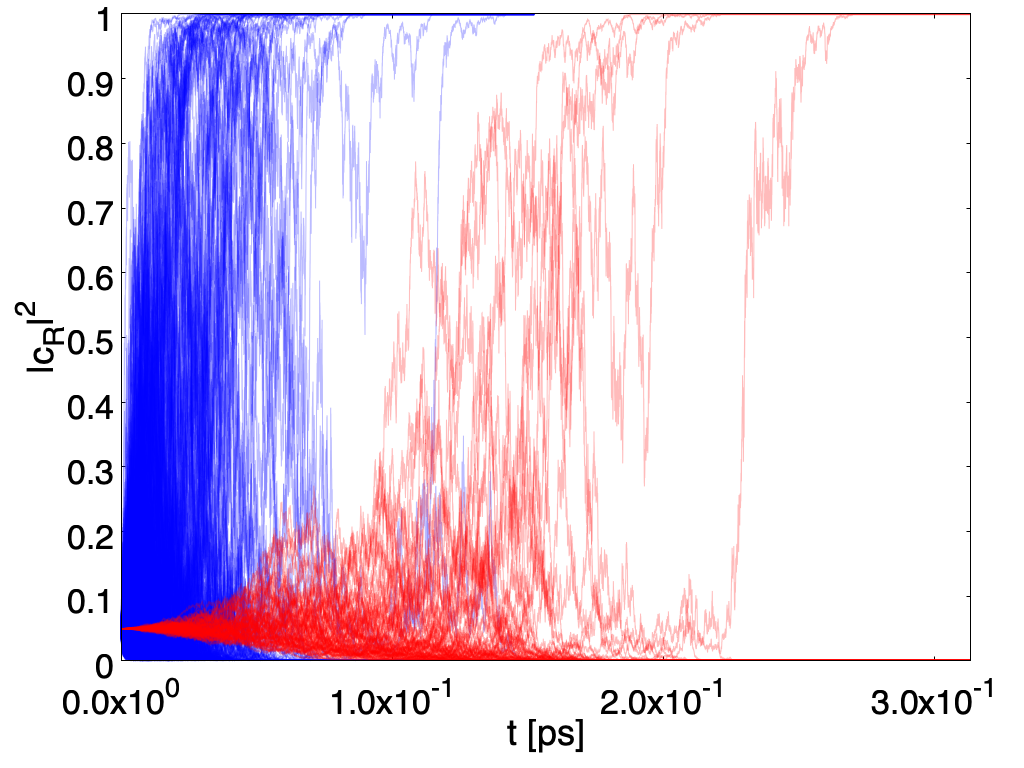}\label{twoDetector_m1_d}}
\caption{$N=1000$ paths of the probability of having the single photon state in the $R$ sector ($|c_R|^2$) as function of $t$ in case of two detectors with $\gamma = \gamma_1$. The time scale in ns$=10^{-9}$ s. For all the figures, in blue are the paths for the case without the von Neumann activation delay for the detectors, in red are the paths for $T=10^{-4}$ ns.}
\label{twoDetector_m1}
\end{figure}

\section{Numerical results}
\label{sec:numres}

The original  choice of parameters, for a process of localisation in position space, in~\cite{Ghirardi:1985mt} was
\begin{equation}
\lambda_{micro} = 10^{-16} {\rm s}^{-1} \qquad \alpha^{-1/2} = 10^{-5} {\rm cm}
\end{equation}

Recently, the experimental search of spontaneous X-ray emission~\cite{Piscicchia_2017} has possibly excluded this set of paramenters. 
 A possible choice compatible with~\cite{Piscicchia_2017} is 
\begin{equation}
\lambda_{micro} = 10^{-17} {\rm s}^{-1} \qquad \alpha^{-1/2} = 10^{-4} {\rm cm} , 
\end{equation}
that, according to eq.~(\ref{eq:betamu}), corresponds to 
\begin{equation}
\gamma_{micro} = {\frac{1}{2}} 10^{-9} {\rm cm}^{-2} {\rm s}^{-1}
\end{equation}

\begin{figure}[htb]
\centering
\subfloat[][Starting condition: $|c_R|^2 = \frac{1}{2}$.]{\includegraphics[width=0.45\textwidth]{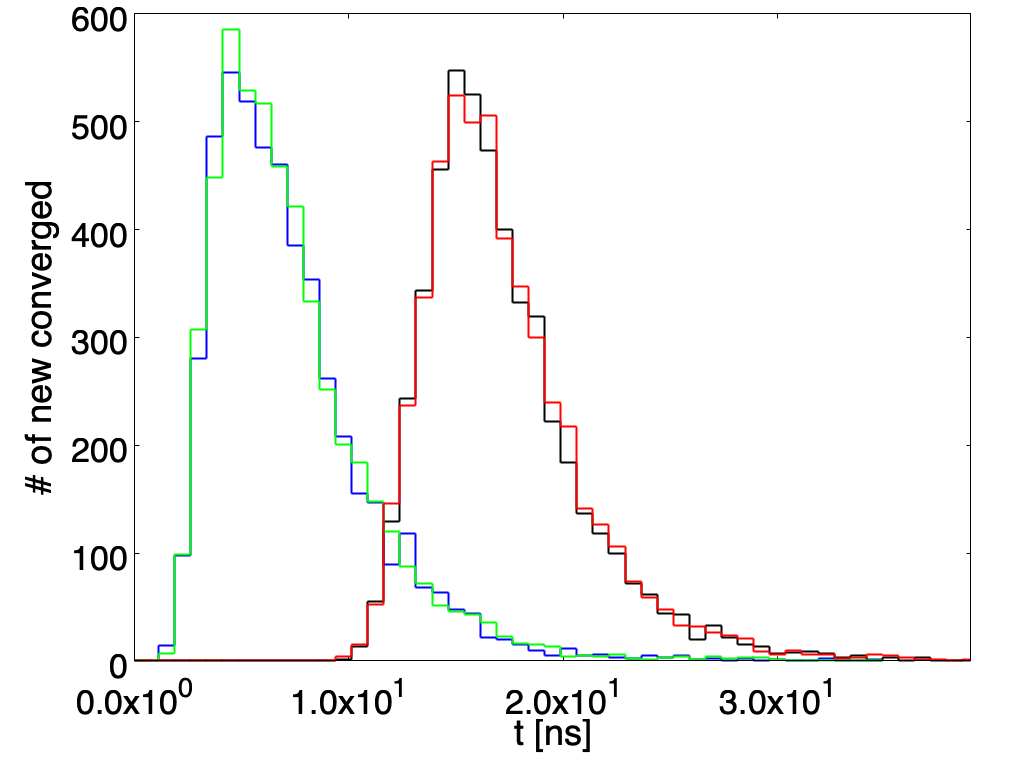}\label{oneDetector_m2_a}}
\subfloat[][Starting condition: $|c_R|^2 = \frac{2}{3}$.]{\includegraphics[width=0.45\textwidth]{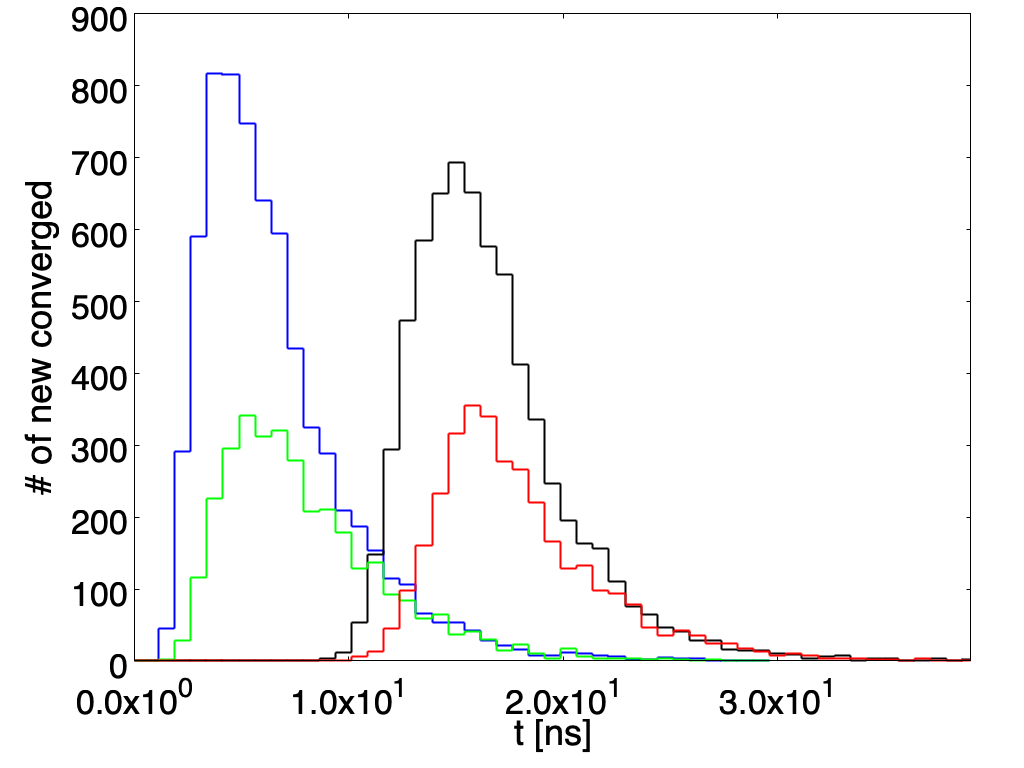}\label{oneDetector_m2_b}} \\
\subfloat[][Starting condition: $|c_R|^2 = \frac{4}{5}$.]{\includegraphics[width=0.45\textwidth]{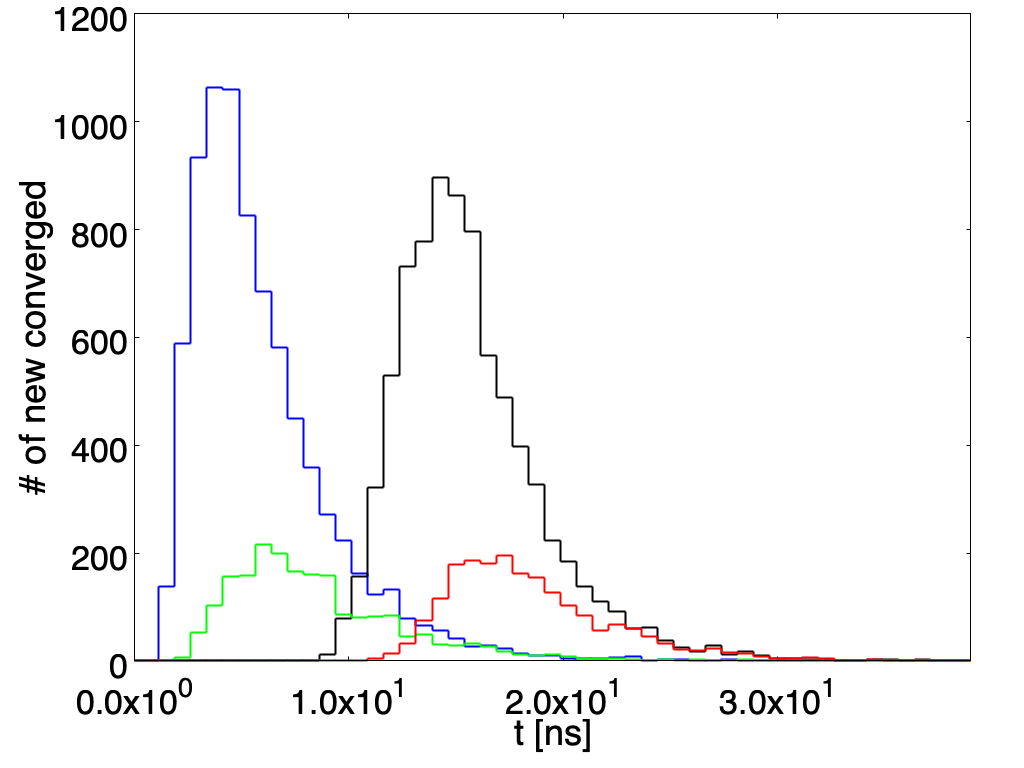}\label{oneDetector_m2_c}}
\subfloat[][Starting condition: $|c_R|^2 = \frac{1}{20}$.]{\includegraphics[width=0.45\textwidth]{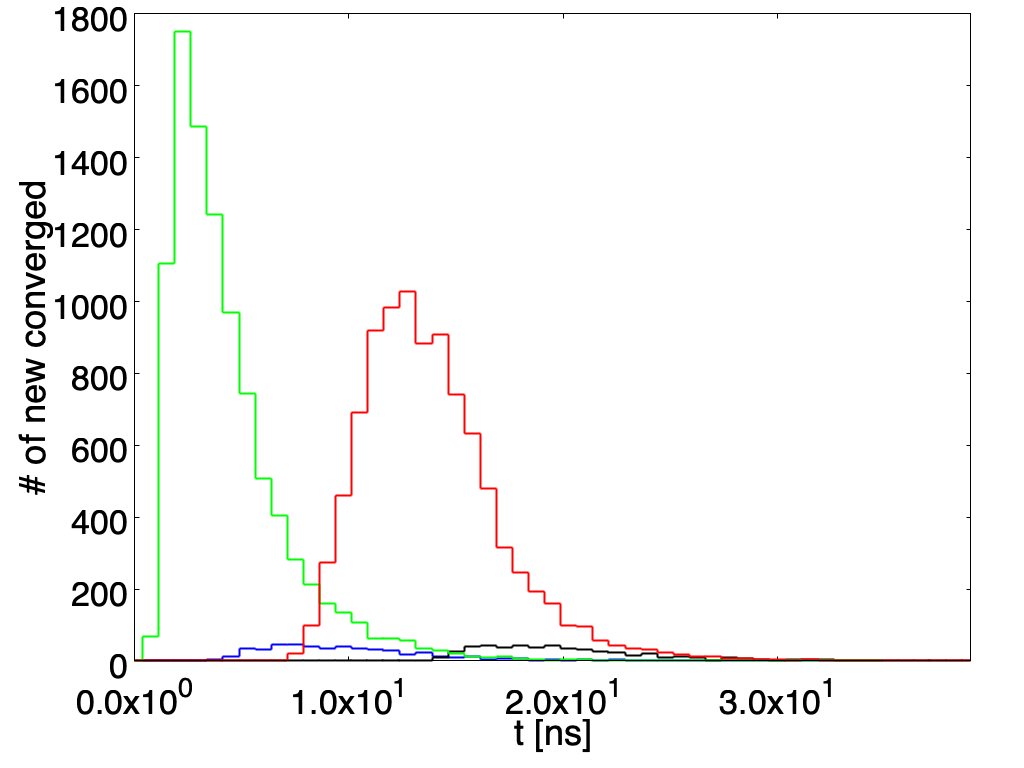}\label{oneDetector_m2_d}}
\caption{The number of paths that, at each time bin, reached $\vert c_R \vert^2 \geq 1-\varepsilon$ (blue and black histograms for  von Neumann activation   delay for the detector switched off/on, respetively) 
or $\vert c_R \vert^2 \leq \varepsilon$ (green and red histograms for  von Neumann activation delay switched off/on, respectively), with $\varepsilon = 1/N$, $N$ being the number of paths considered in the statistical sample ($10^3$ in this case), for  a single detector and $\gamma = \gamma_2$. The integral of any single histogram represents the total number of paths arrived at convergence (Born Rule). }
\label{diff_cumul_oneDetector_m2}
\end{figure}

For an object of approximately $1$ cm$^3$   the number of elementary constituents is assumed to be $10^{23}$. 
Consequently, if the dimension of the object is reduced to $1$ mm$^3$ one has approximately $10^{20}$ constituents.
From these considerations, in the first case one has 
\[
\gamma_{\text{macro,1}} \equiv \gamma_1 =10^{23}\gamma_{\text{micro}} =  \frac{1}{2}10^{14} \text{ cm}^{-2}\text{s}^{-1},
\]
and in the second case
\[
\gamma_{\text{macro,2}} \equiv \gamma_2 =10^{20}\gamma_{\text{micro}} =  \frac{1}{2}10^{11} \text{ cm}^{-2}\text{s}^{-1}. 
\]
Correspondingly, it is assumed that the parameter $a$ describilng the ``pointer" shift in position is $a = 1$~cm in the first case and 
$a = 1$~mm in the second one. 
 
\begin{figure}[htb]
\centering
\subfloat[][Starting condition: $|c_R|^2 = \frac{1}{2}$.]{\includegraphics[width=0.45\textwidth]{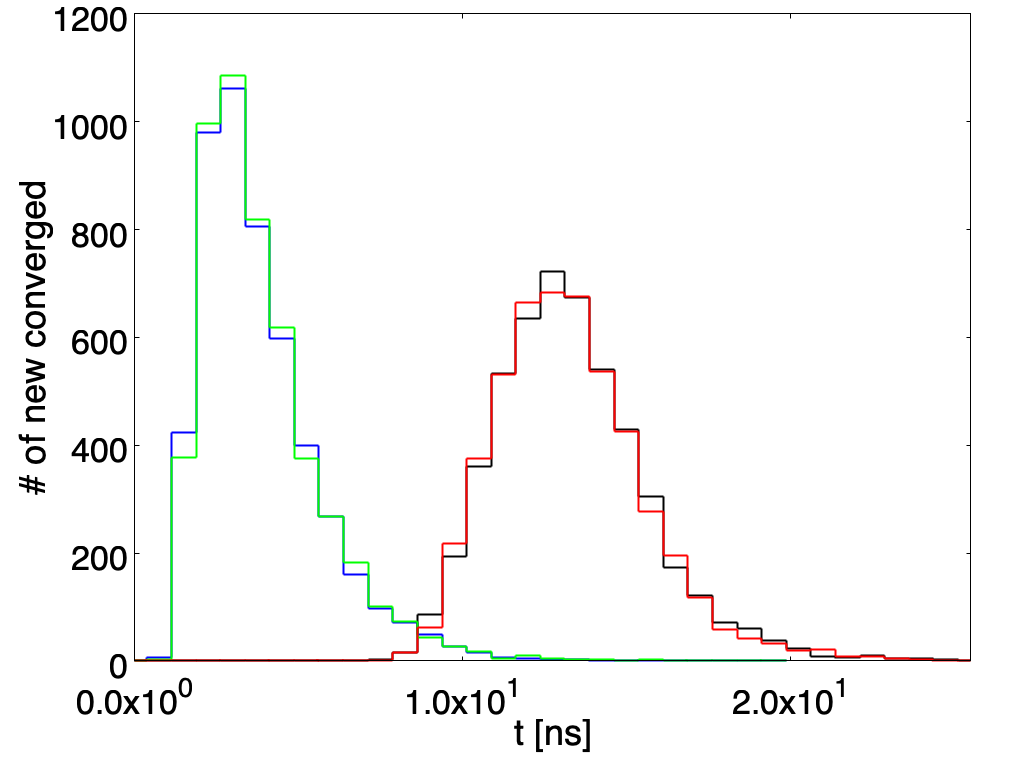}\label{twoDetector_m2_a}}
\subfloat[][Starting condition: $|c_R|^2 = \frac{2}{3}$.]{\includegraphics[width=0.45\textwidth]{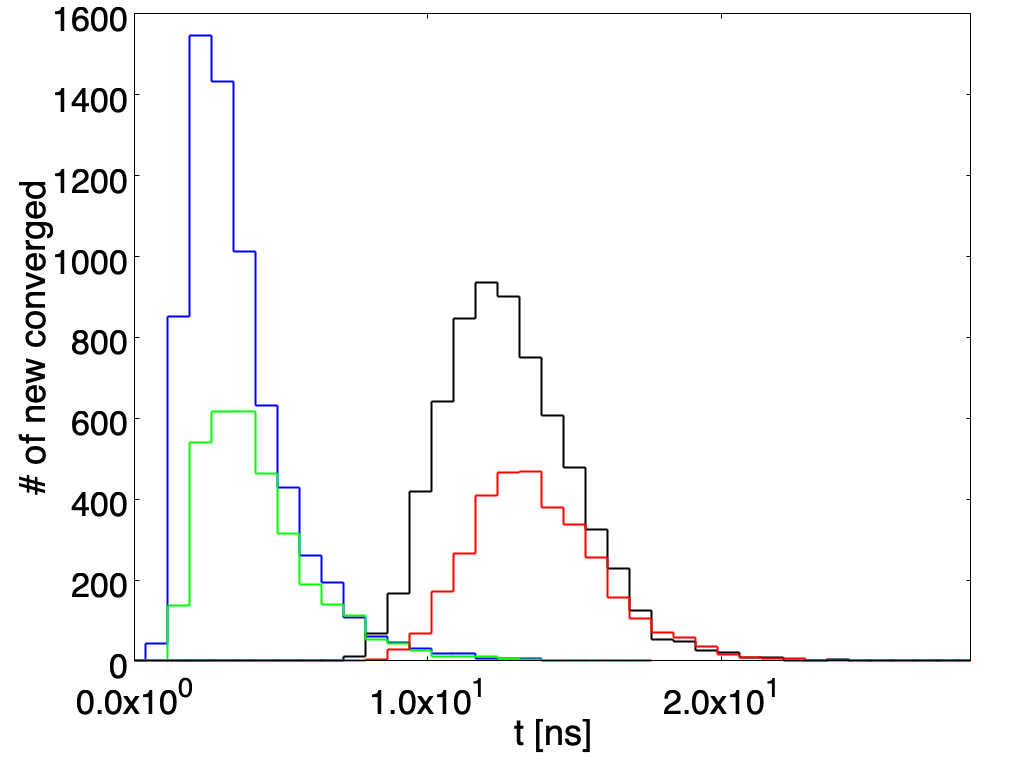}\label{twoDetector_m2_b}} \\
\subfloat[][Starting condition: $|c_R|^2 = \frac{4}{5}$.]{\includegraphics[width=0.45\textwidth]{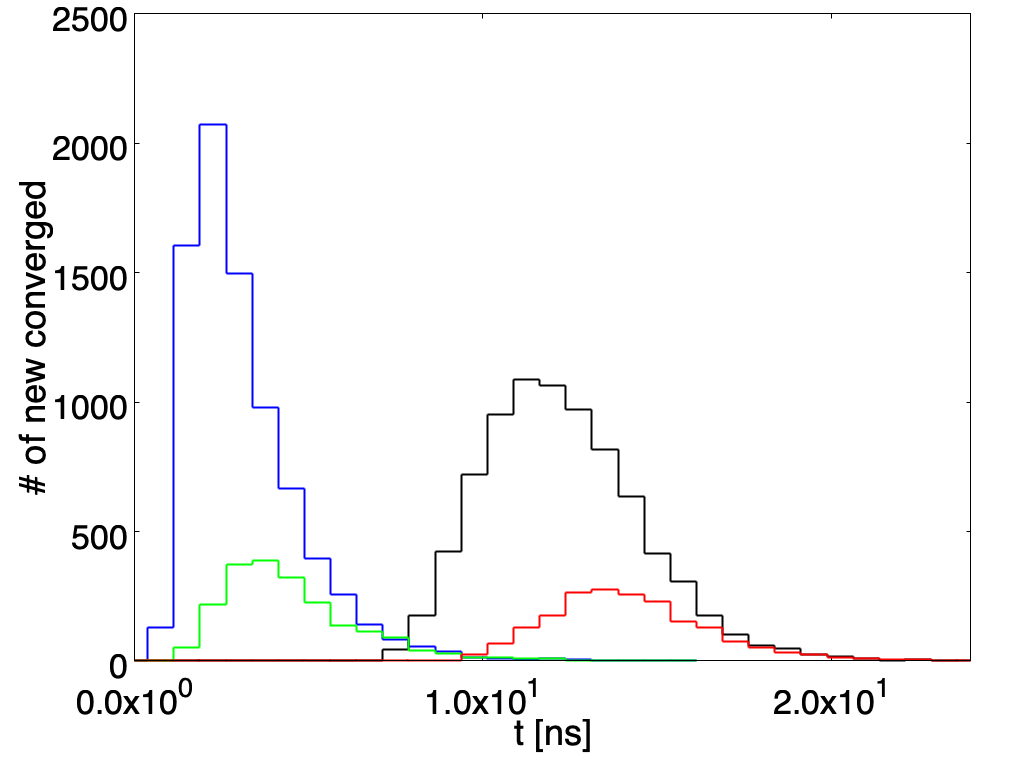}\label{twoDetector_m2_c}}
\subfloat[][Starting condition: $|c_R|^2 = \frac{1}{20}$.]{\includegraphics[width=0.45\textwidth]{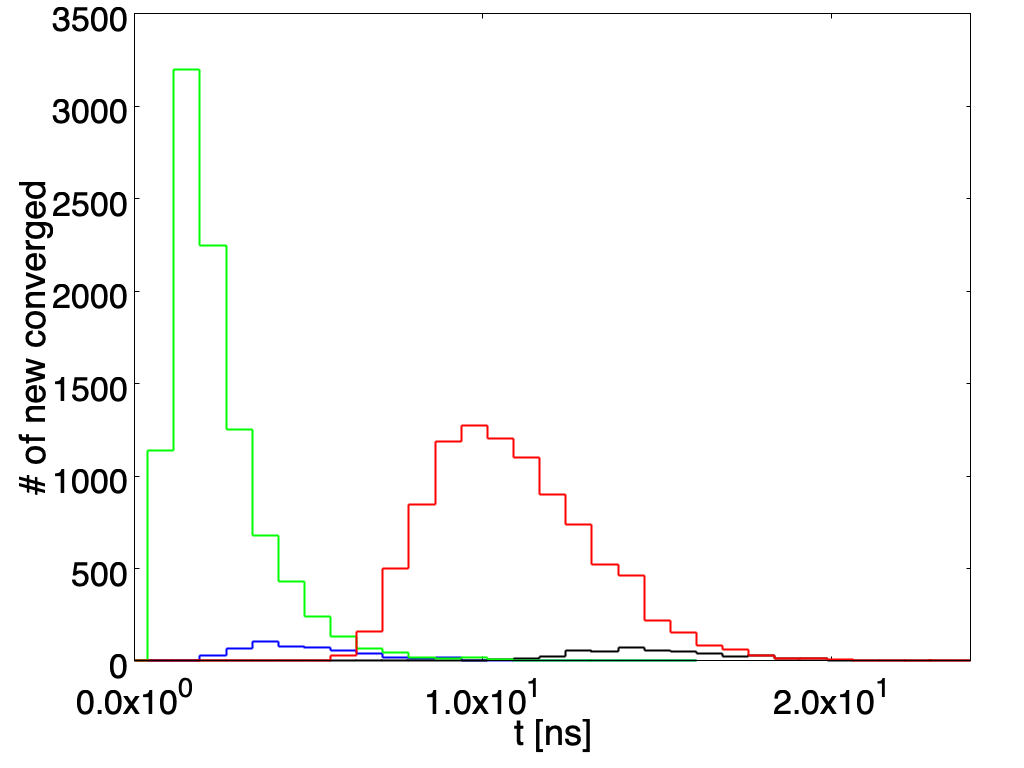}\label{twoDetector_m2_d}}
\caption{The number of paths that, at each time bin, reached $\vert c_R \vert^2 \geq 1-\varepsilon$ (blue and black histograms for  von Neumann activation   delay for the detector switched off/on, respetively) 
or $\vert c_R \vert^2 \leq \varepsilon$ (green and red histograms for  von Neumann activation delay switched off/on, respectively), with $\varepsilon = 1/N$, $N$ being the number of paths considered in the statistical sample ($10^3$ in this case), for the two detectors case and $\gamma = \gamma_2$. The integral of any single histogram represents the total number of paths arrived at convergence (Born Rule).  }
\label{diff_cumul_twoDetector_m2}
\end{figure}

Figure~\ref{oneDetector_m1} shows a sample of $10^3$ paths followed by $\vert c_R (t) \vert^2$ for the starting conditions 
$\vert c_R \vert^2 = 1/2, 2/3, 4/5$ and $1/20$ for the case of a single detector with $\gamma = \gamma_1$. The blue/red paths correspond to the effect of the stochastic process alone and to the combined effect of stochastic process and  von Neumann activation with $T=10^{-4}$~ns, respectively. The paths tend to converge to $\vert c_R \vert^2 = 1$ or $\vert c_R \vert^2 = 0$ with probability 
given by the starting condition (Born rule). For the stochastic process alone, the time scale at which the paths approach convergence is of the order of $5 \times 10^{-5}$~ns, while in the presence of  von Neumann activation the convergence is more distributed in time and becomes clear for $t \simeq T$, as expected.

Figure~\ref{oneDetector_m2} is the same as Figure~\ref{oneDetector_m1}, but for $\gamma = \gamma_2$ and $T = 5$~ns. 
Now the time scale at which the paths approach convergence is of the order of $3$~ns (no von Neumann activation). The lengthening of the convergence time is due to the fact that in this case the detector is smaller and hence fewer constituents contribute to the spontaneous collapse. 

\begin{figure}[htb]
\begin{center}
\includegraphics[width=0.7\textwidth]{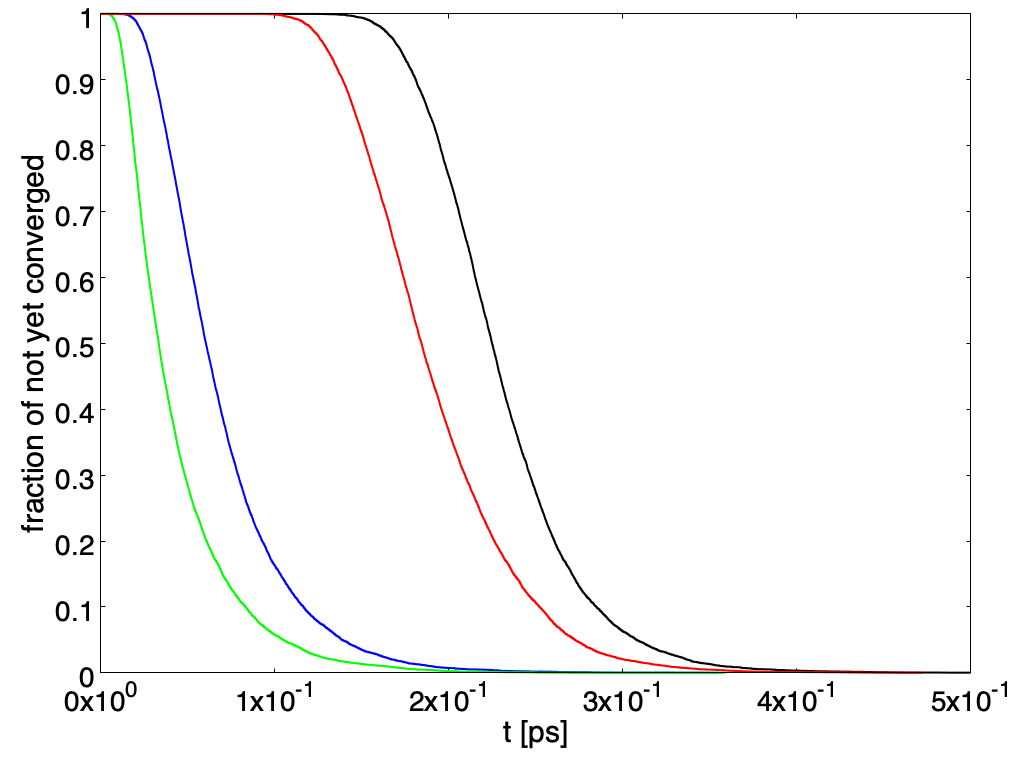}
\end{center}
\caption{Persistence of the superposition. Fraction of paths with $\varepsilon \leq \vert c_R \vert^2 \leq 1-\varepsilon$. The green and blue curves correspond to the starting conditions $ \vert c_R \vert^2 = 1/20 $ and $1/2$, respectively, without the von Neumann activation. The red and black curves represent the same situation with  von Neumann activation. The setup with $\gamma = \gamma_1$ and a single detector has been chosen. }
\label{fig:notConv}
\end{figure}

Figure~\ref{twoDetector_m1} is the same as Figure~\ref{oneDetector_m1}, but for the case of two detectors. As can be seen, the convergence is more rapid than in the case of a single detector. The larger effectiveness of the localization process can be traced back to the statistical properties of the stochastic process $\diff C$, whose variance is two times the variance of the individual stochastic processes $\diff B$. 

In all these cases $T$ has been chosen larger than the typical convergence time scale; for $T$ less then, or similar to, the convergence time scale the stochastic process is not affected in a significant way.

In order to better quantify the convergence properties of the processes, Figure~\ref{diff_cumul_oneDetector_m2} shows the number of paths that, at each time bin, reached $\vert c_R \vert^2 \geq 1-\varepsilon$ (blue and black histograms for   von Neumann activation switched off/on, respetively) 
or $\vert c_R \vert^2 \leq \varepsilon$ (green and red histograms for   von Neumann activation switched off/on, respetively), with $\varepsilon = 1/N$, $N$ being the number of paths considered in the statistical sample, for the case of a single detector and $\gamma = \gamma_2$. The integral of any single histogram represents the total number of paths arrived at convergence (Born Rule). As can be noticed by comparing same color histograms, their shape and peak position depend on the starting conditions. For instance, looking at the blue histograms (paths converging to $\vert c_R \vert^2 = 1$ ), the peak present for the starting condition $\vert c_R \vert^2 = 4/5$ tends to flatten as  the initial value for $\vert c_R \vert^2 $ is reduced and its position shifts from about 5 to about 8~ns for the starting condition 
$\vert c_R \vert^2  = 1/20$. Qualitatively similar comments hold also for Figure~\ref{diff_cumul_twoDetector_m2}, where the case of two detectors is shown.

Figure~\ref{fig:notConv} shows the persistence of the superposition of $\ket{\psi_{R,L}}$ states. The fraction of paths with $\varepsilon \leq \vert c_R \vert^2 \leq 1-\varepsilon$ is represented as a function of time. The green and blue curves correspond to the starting conditions $ \vert c_R \vert^2 = 1/20 $ and $1/2$, respectively, without  von Neumann activation. The red and black curves represent the same situation with  von Neumann activation. The setup with $\gamma = \gamma_1$ and a single detector has been chosen. As can be seen, both with and without  von Neumann activation the starting condition $ \vert c_R \vert^2 = 1/2 $ corresponds to a longer lasting superposition, while with $ \vert c_R \vert^2 = 1/20 $ the superposition decays more rapidly. A qualitatively similar situation is found for the other setup previously considered.

\begin{figure}[htb]
\begin{center}
\includegraphics[width=0.7\textwidth]{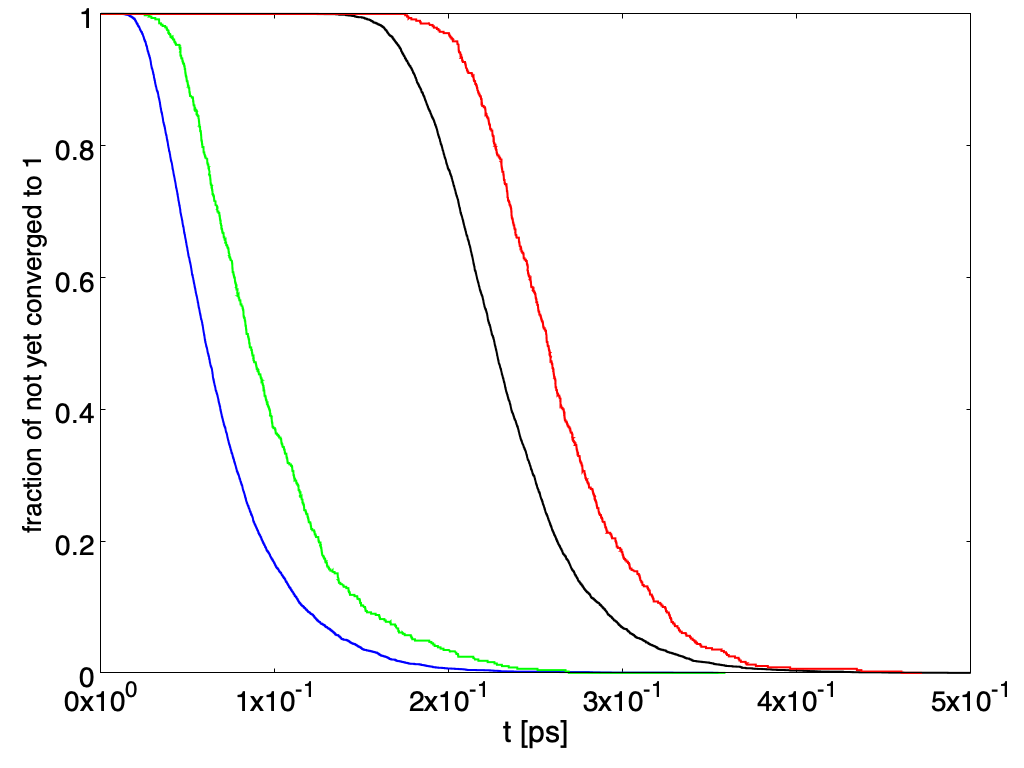}
\end{center}
\caption{Persistence of the superposition. Fraction of paths with $\varepsilon \leq \vert c_R \vert^2 \leq 1-\varepsilon$ as a function of time, for the paths converging to $ \vert c_R \vert^2 = 1$. The blue and green curves correspond to the starting conditions $ \vert c_R \vert^2 = 1/2 $ and $1/20$, respectively, without  von Neumann activation. The black and red curves represent the same situation with  von Neumann activation. The setup with $\gamma = \gamma_1$ and a single detector has been chosen. }
\label{fig:notConvup}
\end{figure}

At last, Figure~\ref{fig:notConvup} shows the persistence of the superposition for the paths converging to $ \vert c_R \vert^2 = 1$. The fraction of paths  converging to $ \vert c_R \vert^2 = 1$ with $\varepsilon \leq \vert c_R \vert^2 \leq 1-\varepsilon$ is represented as a function of time,  the blue and green curves correspond to the starting conditions $ \vert c_R \vert^2 = 1/2 $ and $1/20$, respectively, without  von Neumann activation. The black and red curves represent the same situation with  von Neumann activation. The setup with $\gamma = \gamma_1$ and a single detector has been chosen. As can be seen, both with and without  von Neumann activation, for the paths converging to $ \vert c_R \vert^2 = 1$ the superposition of $\ket{\psi_{R,L}}$ states is more persistent for the starting condition $ \vert c_R \vert^2 = 1/20$ than for $ \vert c_R \vert^2 = 1/2$. 

\section{Conclusions}
\label{sect:concl}
In the present paper a detailed study of  collapse dynamics as provided by GRW theory 
and its continuous realisations in the form of stochastic differential equations describing a 
Brownian-driven motion in Hilbert space has been presented. The possible effect of the finite reaction 
time of the apparatus has also been considered.  It has been assumed that 
a ``pointer" shifts its position by an amount $a$ as a consequence of its measurement; this, of course, 
is not the general case, and what is the ``pointer" has to be established in relation to the real 
physical detection and logging apparatus employed. The features of the 
convergence properties depend on the physical properties of the superposition state, and could be exploited 
to design experiments aiming at pointing out ``GRW" effects.

The two elements determining the time resolution in the experiment outlined in figure~\ref{fig:exp1} or \ref{fig:exp2} are the single photon source and the detectors. We will now briefly consider the challenges each one presents to the possibility of a an actual implementation of the proposed experiment.

A wide variety of single photon sources have been developed over the years, especially thanks to their applied potential in quantum technologies such as quantum communication and quantum computing. Among these sources, the most commonly used in quantum optical experiments fall in two categories: heralded single photon sources and quantum dots. In the first type of source optical nonlinearities are exploited to create photon pairs, followed by the detection of one of the photons and the corresponding projection of the second one on a single photon Fock state \cite{Castelletto2008}. The advantages of these sources are that they can be very brilliant and, more importantly for the present experiment, the extraction efficiency of the photons from the source approaches unity. The main drawback of these sources is that they are probabilistic in nature, and there always exists a non-zero probability that multiple photon pairs are emitted at the same time, thus polluting the single photon state. The typical lifetime of photons generated with parametric sources can be below one picosecond.

In the case of quantum dots, the emission of a stream of single photons is granted by the Fermionic repulsion of electrons confined in quasi 0-dimensional nanostructures \cite{Arakawa2020}, often embedded in a semiconductor substrate. There are two main disadvantages of quantum dots; the first is that they generally operate at temperatures of the order of a few Kelvins, the second is that the extraction efficiency from the source is usually less than 50\%. This second issue is of particular importance for the proposed experiment; indeed  the majority of the photons are generally scattered inside the semiconductor containing the quantum dot, to be quickly absorbed by the substrate. This high probability of absorption before reaching the detectors could result in a change in the collapse dynamics, hindering the experiment. The typical lifetime of photons generated by quantum dots is of the order of tens to hundreds of picoseconds.

Concerning the detectors, the most performant existing single photon detectors at optical and near infrared frequencies are Superconducting Single Photon Detectors  (SSPDs) \cite{Zhang2019}. These devices consist in superconducting wires driven near to the critical current of the superconducting material. If properly designed, the energy of even a single photon absorbed by the wire is sufficient to deposit enough heat to break the superconducting state, thus generating a voltage spike. The time resolution of SSPDs is of the order of a few tens of picoseconds, and their quantum efficiencies are close to unity, generally larger than 95\%.
The voltage spike generated by a detection event in SSPDs is then electronically amplified giving rise to electrical pulses. For a general set-up, we can assume such pulses to be 10 ns in time width and 10 V in amplitude. If the circuit is closed on a 50 Ohm load, each pulse carries a charge of approximatively $10^{10}$ electrons, charge that is provided by capacitive elements within the amplifier circuit (the circuit then needs time to recharge, the so called ``dead time" of the detectors). 

If we assume the electronic pulse following detection events to be the ``pointer", 
and assume that the set of observables to be sharpened is given by mass densities as in the last realisation discussed in ref.~\cite{ShanGao2018}, 
the value of the parameter $\gamma$ would be {\it significantly} smaller than $\gamma_2$, 
resulting in collapse times much longer than 1 ns.  It seems therefore possible to build, with existing technologies, an experiment as that outlined in figure~\ref{fig:exp1} or \ref{fig:exp2} in which the collapse dynamics is longer than the time resolution given by the photon lifetime and the resolution of the detectors (expected to be hundreds of ps at the most). Such an experiment remains however challenging, given in particular the requirement that the almost totality of photons must be succesfully routed to the detectors, to avoid the possibility of collapse due to absorption of the photons from the various objects constituting the experimental set-up. One promising route to minimize this problem might be the use of a fully integrated experiment, in which single photons generated by a quantum dot are not extracted from the semiconductor, but instead emitted in an optical waveguide fabricated in the semiconductor itself. This process can be engineered to have high efficiency thanks to the Purcell effect \cite{Liu2018}. The photons could then be routed toward monolithically integrated SSPDs. Recent experimental results \cite{Schwartz2018} again show that such a goal can be considered within reach of existing photonic technologies.

Albeit the qualitative features of collapse dynamics, as previously shown, do not depend of the value of the parameter $\gamma$, 
collapse times are sensitive to the details of the particular detector employed. A tailored analysis is then required, taking into account all the particular aspects of the experimental setup adopted, and is left to future investigation.


\newpage
\bibliographystyle{unsrt}
\bibliography{CollapseDyn}

\begin{thebibliography}{10}

\bibitem{Schlosshauer:2005}
M.~Schlosshauer.
\newblock Decoherence, the measurement problem, and interpretations of quantum
  mechanics.
\newblock {\em Rev. Mod. Phys.}, 76:1267--1305, Feb 2005.

\bibitem{Ghirardi:1985mt}
G.~C. Ghirardi, A.~Rimini, and T.~Weber.
\newblock {Unified dynamics for microscopic and macroscopic systems}.
\newblock {\em Phys. Rev.}, D34:470, 1986.

\bibitem{ShanGao2018}
G.~C. Ghirardi, O.~Nicrosini, and A.~Rimini.
\newblock What really matters in hilbert-space stochastic processes.
\newblock In Shan Gao, editor, {\em Collapse of the Wave Function: Models,
  Ontology, Origin, and Implications}, chapter~2, pages 12--22. Cambridge
  University Press, 2018.

\bibitem{Wechsler:2020}
S.~Wechsler.
\newblock In praise and in criticism of the model of continuous spontaneous
  localization of the wave-function.
\newblock {\em arXiv: Quantum Physics}, 2020.

\bibitem{Ghirardi:1989cn}
G.~C. Ghirardi, P.~M. Pearle, and A.~Rimini.
\newblock {Markov processes in Hilbert space and continuous spontaneous
  localization of systems of identical particles}.
\newblock {\em Phys. Rev.}, A42:78--79, 1990.

\bibitem{Bassi:2012bg}
A.~Bassi, K.~Lochan, S.~Satin, T.~P. Singh, and H.~Ulbricht.
\newblock {Models of Wave-function Collapse, Underlying Theories, and
  Experimental Tests}.
\newblock {\em Rev. Mod. Phys.}, 85:471--527, 2013.

\bibitem{Diosi:1988}
L.~Di\'osi.
\newblock Continuous quantum measurement and it$\hat{\text{o}}$ formalism.
\newblock {\em Physics Letters A}, 129(8):419--423, 1988.

\bibitem{Gisin:1984}
N.~Gisin.
\newblock Quantum measurements and stochastic processes.
\newblock {\em Phys. Rev. Lett.}, 52:1657--1660, May 1984.

\bibitem{Adler_2001}
S.~L. Adler, D.~C. Brody, T.~A. Brun, and L.~P. Hughston.
\newblock Martingale models for quantum state reduction.
\newblock {\em Journal of Physics A: Mathematical and General},
  34(42):8795–8820, Oct 2001.

\bibitem{Adler_2002}
S.~L. Adler.
\newblock Environmental influence on the measurement process in stochastic
  reduction models.
\newblock {\em Journal of Physics A: Mathematical and General},
  35(4):841–858, Jan 2002.

\bibitem{Adler_2003}
S.~L. Adler.
\newblock Weisskopf-wigner decay theory for the energy-driven stochastic
  schrödinger equation.
\newblock {\em Physical Review D}, 67(2), Jan 2003.

\bibitem{Brody_2002}
D.~C. Brody and L.~P. Hughston.
\newblock Efficient simulation of quantum state reduction.
\newblock {\em Journal of Mathematical Physics}, 43(11):5254–5261, Nov 2002.

\bibitem{Di_si_1988}
L.~Diósi.
\newblock Continuous quantum measurement and itô formalism.
\newblock {\em Physics Letters A}, 129(8-9):419–423, Jun 1988.

\bibitem{Gisin_e1}
N.~Gisin.
\newblock Stochastic quantum dynamics and relativity.
\newblock {\em Helv. Phys. Acta}, 62:363–371, 1989.

\bibitem{Hughston_e1}
L.~P. Hughston.
\newblock Geometry of stochastic state reduction.
\newblock {\em Proc. Roy. Soc. Lond. A}, 452:953–979, 1996.

\bibitem{Penrose_e1}
R.~Penrose.
\newblock On gravity’s role in quantum state reduction.
\newblock {\em General Relativity and Gravitation}, 28:581–600, 1996.

\bibitem{Percival_e1}
I.~C. Percival.
\newblock Primary state diffusion.
\newblock {\em Proc. R. Soc. Lond. A}, 447:189–209, 1994.

\bibitem{NicrosiniRimini:1990}
{Nicrosini, O. and Rimini, A.}
\newblock {On the relationship between continuous and discontinuous stochastic
  processes in Hilbert space}.
\newblock {\em Found. Phys.}, 20:1317--1327, 1990.

\bibitem{Mello2010}
P.~Mello and L.~Johansen.
\newblock Measurements in quantum mechanics and von neumann's model.
\newblock {\em AIP Conference Proceedings}, 1319, 12 2010.

\bibitem{Adler:2020}
S.~L. Adler, A.~Bassi, and L.~Ferialdi.
\newblock Minimum measurement time: lower bound on the frequency cutoff for
  collapse models.
\newblock {\em Journal of Physics A: Mathematical and Theoretical},
  53(21):215302, May 2020.

\bibitem{Piscicchia_2017}
K.~Piscicchia, A.~Bassi, C.~Curceanu, R.~Grande, S.~Donadi, B.~Hiesmayr, and
  A.~Pichler.
\newblock Csl collapse model mapped with the spontaneous radiation.
\newblock {\em Entropy}, 19(7):319, Jun 2017.

\bibitem{Castelletto2008}
S.~A. Castelletto and R.~E. Scholten.
\newblock Heralded single photon sources: a route towards quantum communication
  technology and photon standards.
\newblock {\em The European Physical Journal Applied Physics}, 41(3):181--194,
  March 2008.

\bibitem{Arakawa2020}
A.~Yasuhiko and J.~H. Mark.
\newblock Progress in quantum-dot single photon sources for quantum information
  technologies: A broad spectrum overview.
\newblock {\em Applied Physics Reviews}, 7(2):021309, June 2020.

\bibitem{Zhang2019}
H.~Zhang, L.~Xiao, B.~Luo, J.~Guo, L.~Zhang, and J.~Xie.
\newblock The potential and challenges of time-resolved single-photon detection
  based on current-carrying superconducting nanowires.
\newblock {\em Journal of Physics D: Applied Physics}, 53(1):013001, October
  2019.

\bibitem{Liu2018}
F.~Liu, A.~J. Brash, J.~O'Hara, L.~M. P.~P. Martins, C.~L. Phillips, R.~J.
  Coles, B.~Royall, E.~Clarke, C.~Bentham, N.~Prtljaga, I.~E. Itskevich, L.~R.
  Wilson, M.~S. Skolnick, and A.~M. Fox.
\newblock High purcell factor generation of indistinguishable on-chip single
  photons.
\newblock {\em Nature Nanotechnology}, 13(9):835--840, July 2018.

\bibitem{Schwartz2018}
M.~Schwartz, E.~Schmidt, U.~Rengstl, F.~Hornung, S.~Hepp, S.~L. Portalupi,
  K.~llin, M.~Jetter, M.~Siegel, and P.~Michler.
\newblock Fully on-chip single-photon hanbury-brown and twiss experiment on a
  monolithic semiconductor{\textendash}superconductor platform.
\newblock {\em Nano Letters}, 18(11):6892--6897, October 2018.

\end{thebibliography}

%
%

%
%



\end{document}